\documentclass{aa} 
\usepackage{graphicx}
\usepackage{txfonts}
\usepackage{subfigure}
\usepackage{natbib}

\begin{document}
   \title{The molecular gas content of the Pipe Nebula}

   \subtitle{I. Direct evidence of outflow-generated turbulence in B59?}

\author{A. Duarte-Cabral\inst{1,2,3}
  \and A. Chrysostomou\inst{4}  
  \and N. Peretto\inst{5}
  \and G. A. Fuller\inst{3}
  \and B. Matthews\inst{6,7}  
  \and G. Schieven\inst{6,7}
  \and G. R. Davis\inst{4}  
  }

\offprints{Ana Duarte Cabral, \email{Ana.Cabral@obs.u-bordeaux1.fr}}

 \institute{Univ. Bordeaux, LAB, UMR 5804, F-33270, Floirac, France.
 	 \and CNRS, LAB, UMR 5804, F-33270, Floirac, France
        \and Jodrell Bank Centre for Astrophysics, School of Physics and Astronomy, University of Manchester, Oxford Road, Manchester, M13 9PL, U.K.
        \and Joint Astronomy Centre, 660 North AÕohoku Place, Hilo, HI 96720, USA
      	\and Laboratoire AIM, CEA/DSM-CNRS-Universit\'e Paris Diderot, IRFU/Service d'Astrophysique, 
  C.E. Saclay, Orme de merisiers, 91191 Gif-sur-Yvette, France
        \and Herzberg Institute of Astrophysics, National Research Council Canada, 5071 West Saanich Road., Victoria, BC, Canada, V9E 2E7, Canada
        \and University of Victoria, Finnerty Road, Victoria, BC, V8W 3P6 Canada
        }

\date{Received 18 March 2012; accepted 12 May 2012}

 
  \abstract
  {Star forming regions may share many characteristics, but the specific
    interplay between gravity, magnetic fields, large-scale dynamics, and
    protostellar feedback will have an impact on the star formation history of
    each region. The importance of feedback from outflows is a particular
    subject to debate, as we are yet to understand the details of their impact
    on clouds and star formation.}
  {The Pipe Nebula is a nearby molecular cloud hosting the B59 region as its
    only active star-forming clump. 
    This paper focuses on the global dynamics of B59, its temperature
    structure, and its outflowing gas, with the goal of revealing the local
    and global impact of the protostellar outflows.}
  {Using HARP at the JCMT, we have mapped the B59 region in the
    $J=3\rightarrow2$ transition of $^{12}$CO to study the kinematics and
    energetics of the outflows, and the same transitions of $^{13}$CO and
    C$^{18}$O to study the overall dynamics of the ambient cloud, the physical
    properties of the gas, and the hierarchical structure of the region.}
  {The B59 region has a total of $\sim$~30~M$_{\odot}$ of cold and quiescent
    material, mostly gravitationally bound, with narrow line widths
    throughout. Such low levels of turbulence in the non-star-forming regions
    within B59 are indicative of the intrinsic initial conditions of the
    cloud. On the other hand, close to the protostars the impact of the
    outflows is observed as a localised increase of both C$^{18}$O line widths from
    $\sim0.3$~km\,s$^{-1}$ to $\sim1$~km\,s$^{-1}$, and $^{13}$CO excitation
    temperatures by $\sim2-3$K. The impact of the outflows is also evident in
    the low column density material which shows signs of being shaped by the
    outflow bow shocks as they pierce their way out of the cloud. Much of this
    structure is readily apparent in a dendrogram analysis of the cloud and
    demonstrates that when decomposing clouds using such techniques a careful
    interpretation of the results is needed.}
  {The low mass of B59 together with its intrinsically quiescent gas and small
    number of protostars, allows the identification of specific regions where
    the outflows from the embedded sources interact the dense gas. Our study
    suggests that outflows are an important mechanism for injecting and
    sustaining supersonic turbulence at sub-parsec size scales. We find that
    less than half of the outflow energy is deposited as turbulent energy of
    the gas, however this turbulent energy is sufficient to slow down the
    collapse of the region.}
   

\titlerunning{Pipe Nebula I. Direct evidence of outflow-generated turbulence in B59}
   \maketitle
%

\section{Introduction}

As stellar nurseries, molecular clouds contain the ingredients for star
formation. However, the physical processes which prevent or trigger star
formation have been the subject of debate over the past decade
\citep[e.g.][]{2007ARA&A..45..565M,2010arXiv1009.3962V,2011IAUS..270..159H}.
The relative importance of the various processes is unlikely to be universal:
while in some regions dynamical effects such as a collision of clouds or
convergence of flows
\citep[e.g.][]{2007ApJ...657..870V,2010A&A...520A..49S,2011A&A...528A..50D}
may be significant, other regions can evolve more quietly, from quiescent
material possibly controlled by the local magnetic field
\citep[e.g.][]{2008ApJ...687..354N,2008A&A...486L..13A}.  The imprints of the
initial conditions specific to a given cloud are potentially preserved in the
properties of the gas and dust of their young proto-clusters although once
star formation begins feedback from the forming stars can mask the signatures
of these initial conditions.

Young protostellar outflows may be responsible for clearing and disrupting
protostellar envelopes, limiting the reservoir of mass which will end up being
accreted by the central protostars
\citep[e.g.][]{2002ApJ...573..699F,2006ApJ...646.1070A,2010ApJ...715.1170A}.
Even though an outflow may carry sufficient energy to unbind all of an
envelope, this energy may well be deposited far from the dense gas, or even
outside the molecular cloud. Knowing exactly where the outflow momentum and
energy are deposited is fundamental for understanding whether outflows are a
source of turbulence capable of providing support against gravity, clearing
circumstellar regions and slowing (or enhancing) star formation. However, the
efficiency with which outflows can drive turbulence is strongly debated
\citep[see review from][]{2007ARA&A..45..565M}. The analytical model of
\citet[][]{2000ApJ...545..364M} estimates a typical outflow injection scale,
the scale at which the momentum is most efficiently deposited, of a few tenths
of a parsec.  Numerical models
\citep[e.g.][]{2007ApJ...662..395N,2009ApJ...695.1376C,2010ApJ...722..145C}
confirm that outflows can drive supersonic turbulence at sub-outflow scales
and produce smooth velocity structures on outflow-scales that sweep up and
dissipate smaller structures, even though outflows are not the primary source
of turbulence at all scales in molecular clouds.  The effect of the outflows
is to produce an overall flatter density spectrum, which limits the
fragmentation and infall of material onto the final protostars.

Observationally, the evidence suggests that outflows can provide barely enough
kinetic energy to support entire clouds against gravity
\citep[e.g.][]{2010ApJ...715.1170A,2009A&A...499..175M}, even though they may
be able to maintain the supersonic turbulence at sub-parsec scales
\citep[e.g.][]{2011ApJ...737...56N}.

Nevertheless, the importance of outflows on the energy balance of a region is
likely to depend on the physical properties of the region.
However, it is observationally difficult to measure the direct injection of
turbulence due to outflows and assess its impact on star formation.  One
reason is that in dynamic cluster-forming regions, with large-scale motions,
and on-going star formation, the direct effect of outflows can be
easily confused and masked by other dynamical events.\\

Here we present a study of the B59 star-forming region, within the generally
quiescent Pipe Nebula. Low-density clouds such as the Pipe provide the best
opportunities to directly observe the impact of outflows where stars have began
to form. In Sect.~\ref{sec:pipe} we present an overview of the Pipe Nebula and
its active region, B59. In Sect.~\ref{sec:obs} we detail the observations, and
we present the results on the physical properties in
Sect.~\ref{sec:results:phys}, including optical depth and estimates of
excitation temperatures. The outflows of the region are presented in
Sect.~\ref{b59_outflows}. The ambient cloud's dynamics and hierarchical
structure are described in Sect.~\ref{ambient} (the details on the
hierarchical structure of B59, performed using a dendrogram technique, is
described in detail in Appendix~\ref{hierar}). Finally, we discuss the
outflow-dense gas interaction in Sect.~\ref{discussion:out-dense}, and we
outline our conclusions in Sect.~\ref{concl}.


\section{The Pipe Nebula and B59}
\label{sec:pipe}

\begin{figure}[!t]
	\centering
	{\renewcommand{\baselinestretch}{1.1}
	\includegraphics[angle=270,width=0.5\textwidth]{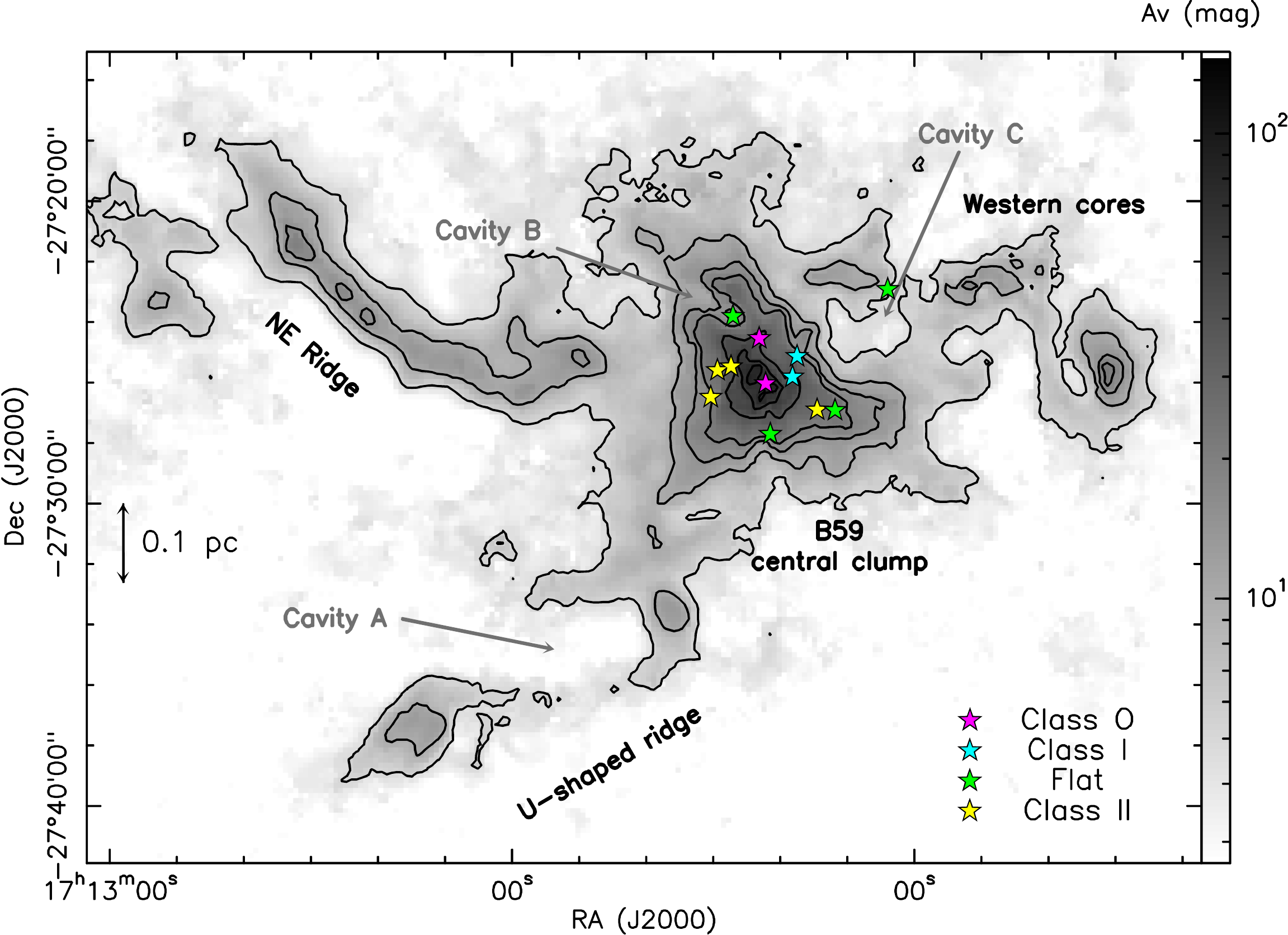}
	\caption[]{\small{Extinction map of B59 in grayscale and contours
            \citep[][]{2009ApJ...704..183R}, showing the known YSOs in the
            region \citep[from][]{2007ApJ...655..364B,2009ApJ...704..292F},
            except for the Class III objects. Contours range from an A$_{v}$
            of 5 to 20 in steps of 5, and then by steps of 20. Note that there
            are only four known Class 0 and I protostars in the region. The
            black labels show designations of regions within B59 as adopted
            throughout the paper. Some cavities are indicated with grey arrows
            and labels, and are discussed further in the text.}}
	\label{fig:b59_ext_sourc}}
\end{figure}

The Pipe Nebula, located at $\sim$
130~pc\footnote{\citet[][]{2007A&A...470..597A} have estimated a distance of
  145~pc, but we adopt 130~pc as it lies within the uncertainties of these
  authors, and it is the distance most commonly adopted in other studies of
  the Pipe Nebula. We note, however, that uncertainties on the distance are of
  the order of 10\% and we are therefore possibly underestimating masses by
  20\% by assuming the nearer distance of 130~pc.}  from the Sun
\citep[][]{2006A&A...454..781L}, is part of the Gould Belt of star-forming
regions.  The global features of this cloud appear similar to other Gould Belt
clouds, with a total mass comparable to the few~$\times$~10$^{4}$~M$_{\odot}$
of the Taurus-Auriga complex, and a filamentary structure with a magnetic
field threading it \citep[][]{2008A&A...486L..13A}. However the Pipe has an
extremely low star formation efficiency of less than 0.1$\%$ compared to other
Gould Belt clouds which have efficiencies of $\sim2-20$\%. In fact, only one
active star forming clump, B59, has been found thus far in the Pipe cloud
\citep[][]{2007ApJ...655..364B,2009ApJ...704..292F,2010ApJ...719..691F}.  This
small group of protostars within B59 are known to power more than one
molecular outflow \citep{1999PASJ...51..871O,2009ApJ...700.1541R}.
Figure~\ref{fig:b59_ext_sourc} shows the extinction map of B59 and identifies
several features and regions which will be referred to in this work.

Several scenarios have been suggested for the formation of B59 as an active
star-forming clump at the edge of the Pipe Nebula. One suggests that B59
formed solely governed by gravity \citep[][]{2009ApJ...704.1735H}. In another
view, the magnetic field plays an important role guiding the collapse of
material towards the main filament
\citep[][]{2008A&A...486L..13A}. \citet[][]{2012A&A...541A..63P} propose that
the formation of B59 at the edge of the Pipe is due to a compression front
(west of the B59, beyond the western cores) from a nearby OB association which
has gathered the gas in a ``snow-plough'' fashion. The effects of
gravitational focusing playing a role later on, with a contraction of the
gathered gas and the formation of stars in the centre of B59.

\begin{figure*}[!t]
	\centering
	{\renewcommand{\baselinestretch}{1.1}
	\includegraphics[angle=270,width=0.98\textwidth]{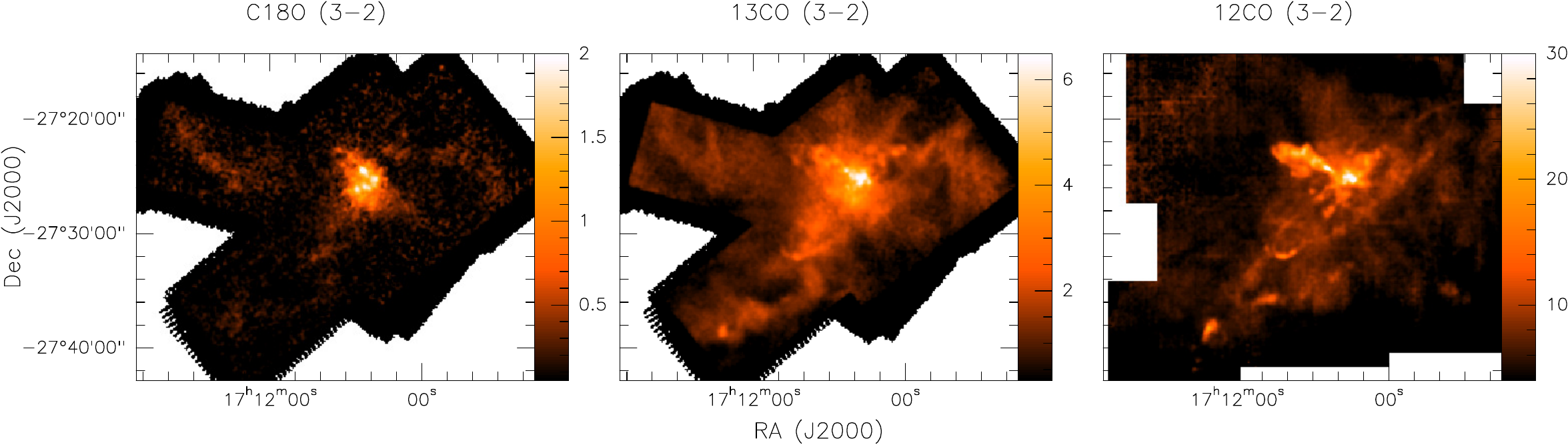}
	\caption{\small{Integrated intensity ($\int {\rm T}_{\rm{A}}^{*} d\,v$
            K\,km\,s$^{-1}$) maps of the three isotopologues observed. For the
            C$^{18}$O (left) and $^{13}$CO (middle), the integration range was
            1 to 5.5~km\,s$^{-1}$. The $^{12}$CO (right) was integrated from
            -5 to 15~km\,s$^{-1}$. Here we can see that the C$^{18}$O follows
            the shape as seen in extinction (Fig.~\ref{fig:b59_ext_sourc}),
            while the $^{13}$CO emission starts to pick up the outflowing
            material, best seen in the $^{12}$CO emission dominated by high
            velocity gas.}}
	\label{fig:b59_integ}}
\end{figure*}

Our previous knowledge of the molecular gas structure in the Pipe Nebula,
however, was limited to the $4^\prime$ resolution maps of the
$J=1\rightarrow0$ transitions of $^{12}$CO, $^{13}$CO and C$^{18}$O by
\citet{1999PASJ...51..871O}. Beyond this, molecular line observations of the
Pipe Nebula consist of single spectra of dense gas tracers such as NH$_{3}$,
CCS, HC$_{5}$N and other early- and late-time molecules
\citep[][]{2008ApJS..174..396R,2010ApJ...723.1665F,2012A&A...537L...9F}, and
C$^{18}$O \citep[][]{2007ApJ...671.1820M} towards the peaks of cores
identified in $1^\prime$ resolution extinction maps
\citep[][]{2006A&A...454..781L,2010ApJ...725.2232R}.


\section{Observations}
\label{sec:obs}

\begin{figure}[!t]
	\centering
	{\renewcommand{\baselinestretch}{1.1}
	\includegraphics[angle=270,width=0.45\textwidth]{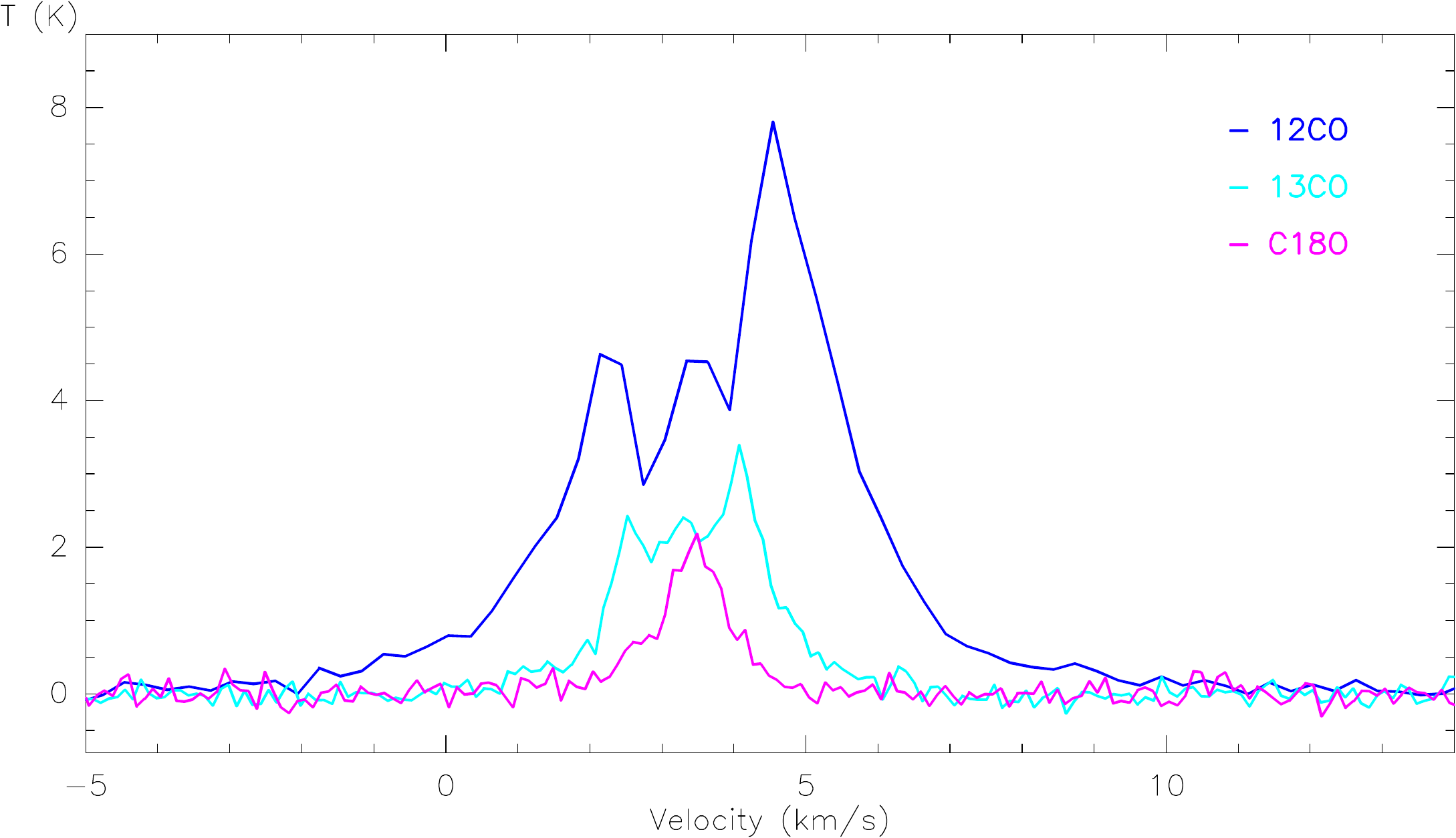}
	\caption[]{\small{Plot of C$^{18}$O (purple line), $^{13}$CO (light
            blue line), with a velocity resolution of 0.1~km\,s$^{-1}$, and
            $^{12}$CO (dark blue line), with a velocity resolution of
            0.3~km\,s$^{-1}$, in B59, at the position of the Class 0 source B11,
            at RA~=~17$^{\rm{h}}$11$^{\rm{m}}$23.0$^{\rm{s}}$,
            Dec~=~-27$^{\circ}$24'32.8'', showing the outflow wings and self
            absorption dip in $^{13}$CO and $^{12}$CO.}}
	\label{fig:b59_spectra}}
\end{figure} 

Using HARP at the JCMT \citep[][]{2009MNRAS.399.1026B} in May and June 2010,
we have mapped $^{13}$CO and C$^{18}$O $J=3\rightarrow2$ (at 330.6 and
329.3~GHz respectively) over the entire B59 star-forming region
($\sim$~0.11~deg$^{2}$). The observations were taken with a sky opacity at
225GHz varying between 0.05 and 0.08. The data were observed as several raster
maps which have an original spatial resolution of 15$^{\prime\prime}$ and a
spectral resolution of 0.05~km\,s$^{-1}$. The r.m.s. noise level on the final
B59 dataset is 0.22~K (in T$_{\mathrm{A}}^{*}$) with 0.25 km\,s$^{-1}$
channels. Complementary $^{12}$CO $J=3\rightarrow2$ (345.8~GHz) data 
were obtained with HARP using JCMT Director's discretionary time. These data
reached an r.m.s. noise level of 0.2~K (in T$_{\mathrm{A}}^{*}$) at
0.5~km\,s$^{-1}$ velocity resolution. All data were corrected for the telescope
main beam efficiency of $\eta_{\mathrm{mb}} = 0.66$
\citep{2009MNRAS.399.1026B,2010MNRAS.401..455C}.

The data reduction was performed using the ORAC-DR pipeline
\citep[][]{2008AN....329..295C}, using the recipe \textsc{reduce science
  narrowline}, with minor modifications tailored for our dataset, with a pixel
size of 7.4$^{\prime\prime}$. This reduction procedure automatically fits and
corrects baselines and removes bad detectors based on the time-series data and
on the r.m.s. noise levels of the final maps. The final reduced maps for each
area were gridded together using Starlink software.

All datacubes were convolved to a 20$^{\prime\prime}$ resolution in order to
suppress some of the high frequency noise, and were masked to pixels where the
peak signal to noise (estimated at each pixel) was higher than 4 in the
smoothed map. This masking procedure was particularly effective for the weaker
C$^{18}$O, but it left virtually unchanged the datacubes for the other two
isotopologues. The reduced integrated intensity maps are presented in
Fig.~\ref{fig:b59_integ}, and Fig.~\ref{fig:b59_spectra} shows an example of
spectra at the position of a protostellar source.


\section{Opacity and temperature structure of B59}
\label{sec:results:phys}

\subsection{CO-to-H$_{2}$ relation}
\label{H2COlinear}

The column density structure of B59 as seen with the dust extinction and
emission is now relatively well known
\citep[][]{2009ApJ...704..183R,2012ApJ...747..149R,2012A&A...541A..63P}.
Unlike the dust and gas emission, the dust extinction is not sensitive to the
temperature. As such, studying the relation between the H$_{2}$ column
densities derived from the dust extinction and the gas emission can give an
insight to optical depth effects, temperature effects and the local abundance
of a molecule with respect to H$_{2}$.
 
\begin{figure*}[!t]
	\centering
	{\renewcommand{\baselinestretch}{1.1}
	\includegraphics[width=0.4\textwidth]{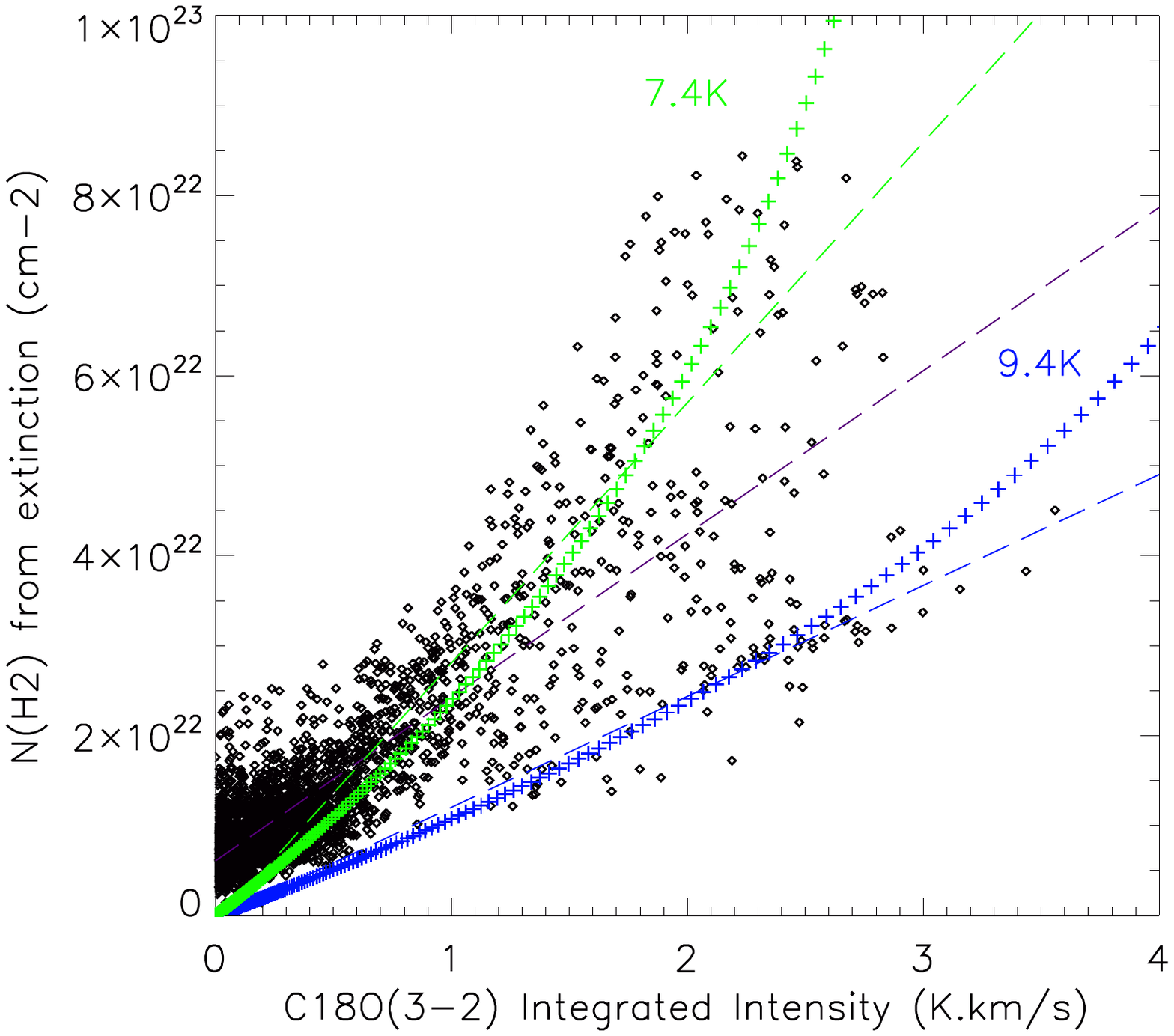}	
	\hspace{1cm}
	\includegraphics[width=0.4\textwidth]{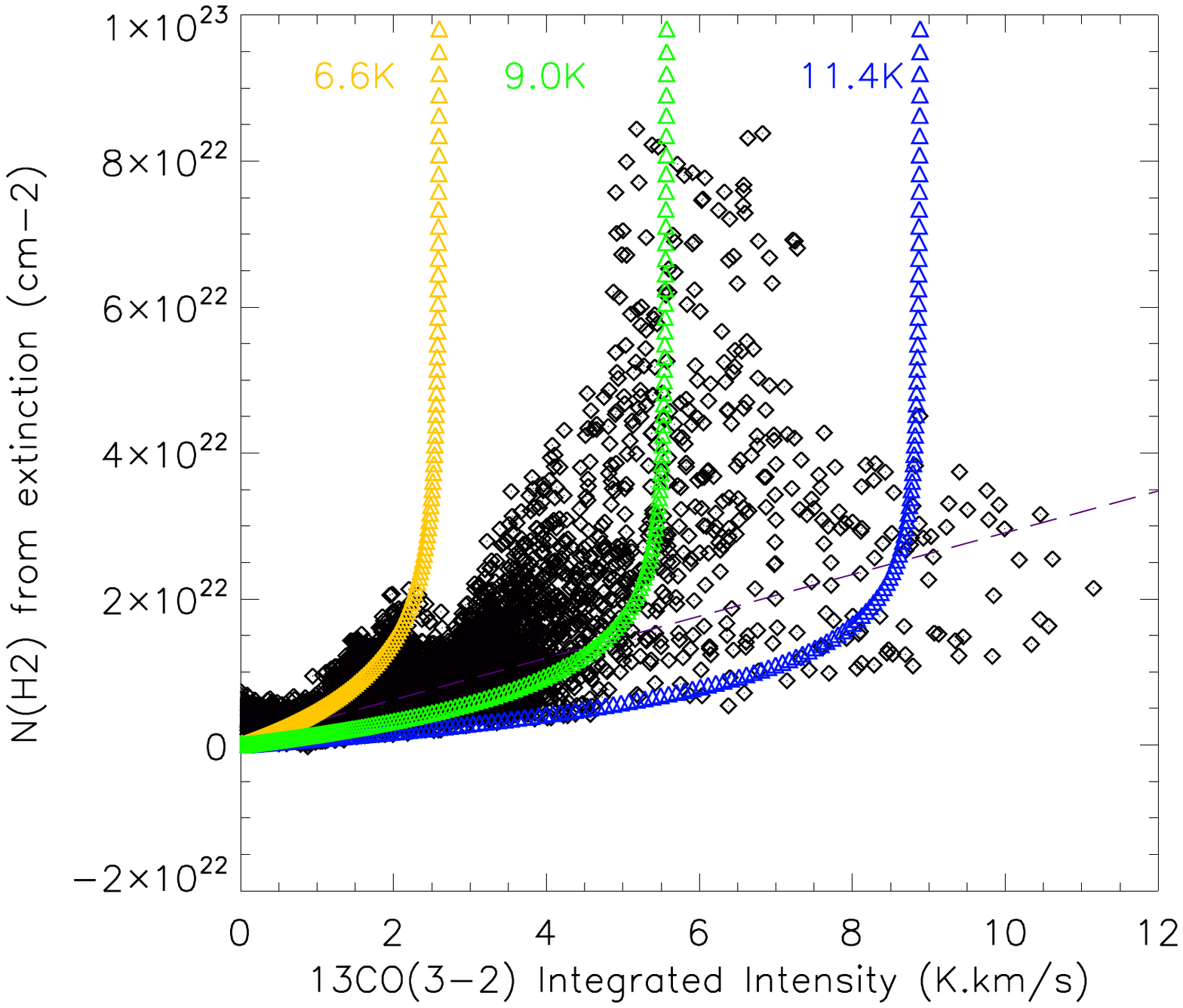}
	\vspace{-3.5cm}
	\caption{\small{Scatter plots of the H$_{2}$ column density derived
            from the extinction map against the C$^{18}$O (left) and $^{13}$CO
            (right) integrated intensity ($\int {\rm T}_{\rm{mb}}^{*} d\,v$
            K\,km\,s$^{-1}$) for each pixel in the map. The purple
            dashed-lines represent the linear fit of the data points, with an
            upper limit constraint on the column density for the case of
            $^{13}$CO. Overlaid on the scatter plots are the predictions of
            RADEX models at different gas temperatures (on the left: 7.4~K in
            green and 9.4~K in blue crosses; on the right: 6.6~K in yellow,
            9.0~K in green and 11.4~K in blue triangles). For the C$^{18}$O
            (left) the green and blue dashed lines represent the linear
            approximations to the respective colour-coded RADEX model.}}
	\label{fig:scatter_13cob59}
	\label{fig:scatter_b59}}
\end{figure*}

Figure~\ref{fig:scatter_13cob59} presents the relation between H$_{2}$ column
densities and the $^{13}$CO and C$^{18}$O emission in B59 as scatter plots.
The H$_{2}$ column density was derived from the extinction map of B59
\citep{2009ApJ...704..183R} assuming $N(\mathrm{H}_{2})/A_{\mathrm{v}} = 9.4
\times 10^{20}$ cm$^{-2}$~mag$^{-1}$
\citep[e.g.][]{1978ApJ...224..132B,2008ApJ...679..481P}. This is plotted
against the $^{13}$CO and C$^{18}$O integrated intensities corrected for the
telescope efficiency. Both the extinction and the molecular line maps have
20$^{\prime\prime}$ resolution and were resampled to a common area and pixel
size, to enable a pixel-by-pixel comparison.

On the figure, the coloured symbols indicate the correlation predicted by the
non-LTE radiative transfer model RADEX \citep[][]{2007A&A...468..627V} for
different kinetic temperatures, using a volume density above the critical
densities of $^{13}$CO and C$^{18}$O (3~-~2). The H$_{2}$ column densities are
estimated assuming a fractional abundance of $^{13}$CO and C$^{18}$O with
respect to H$_{2}$ of $1.4\times10^{-6}$ and $1.7\times10^{-7}$ respectively
\citep[][]{1982ApJ...262..590F}.  These models are only meant to illustrate
the behaviour of the emission at different temperatures, and they do not
represent actual fits to the datapoints.  The curves show the turn-over
between a linear correlation between gas emission and H$_{2}$ column density
and where the lines become optically thick. The linear portion of the curve is
related to the abundance of $^{13}$CO and C$^{18}$O relative to H$_{2}$.

To estimate the masses and column densities directly from the gas emission
(Appendix~\ref{hierar}), we estimate a ``conversion'' factor, $X_{\rm{CO}}$,
on the linear portion of the datapoints as
$X_{\rm{CO}}=N(\mathrm{H}_{2})/I_{\rm{CO}}$
\citep[e.g.][]{2008ApJ...679..481P}, where $N(\mathrm{H}_{2})$ is the H$_{2}$
column density and $I_{\rm{CO}}$ is the integrated intensity of the molecular
transition.

\begin{figure}[!t]
	\centering
	{\renewcommand{\baselinestretch}{1.1}
	\includegraphics[angle=270,width=0.48\textwidth]{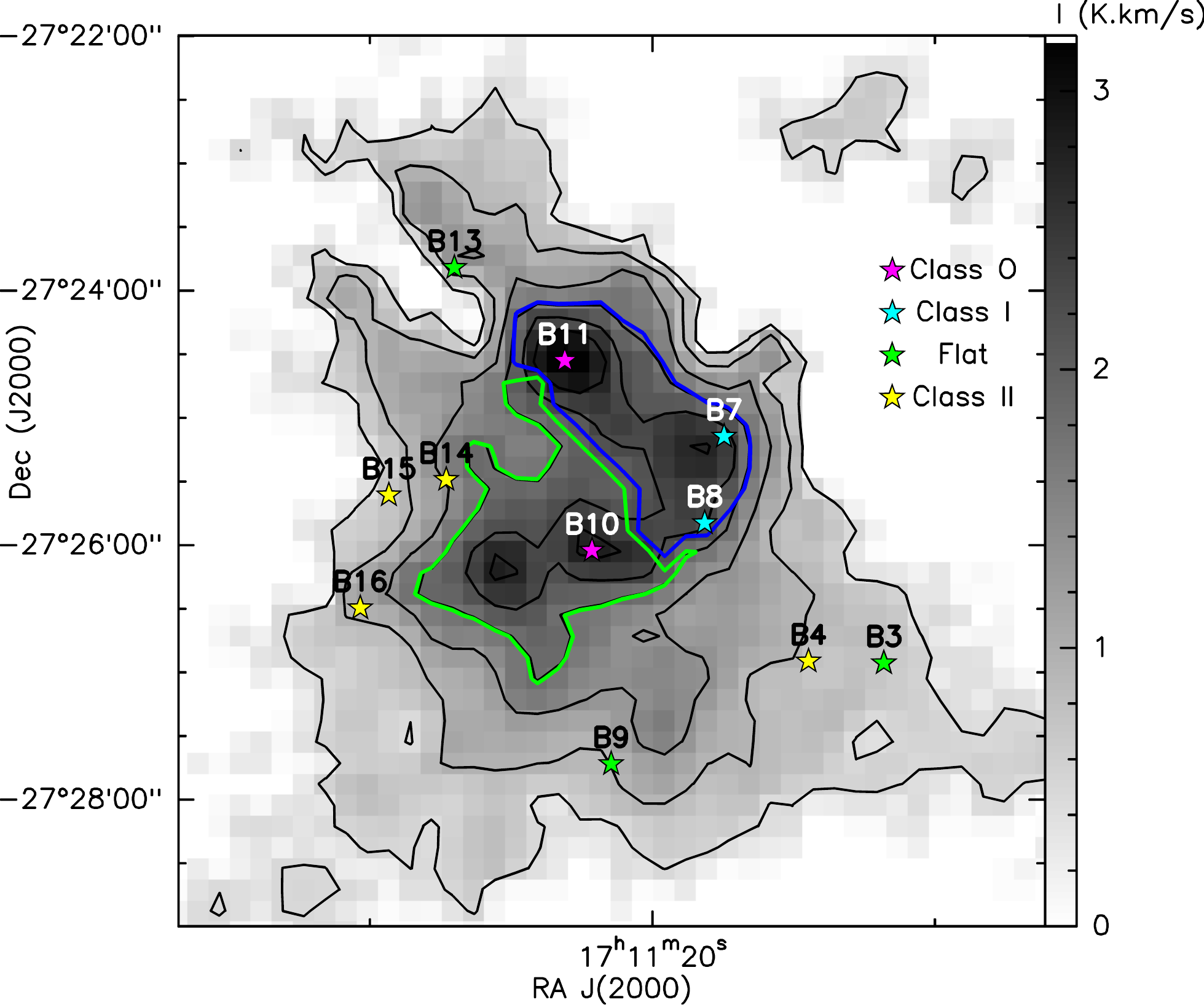}
	\caption{\small{Integrated intensity of C$^{18}$O (3-2) in the central
            region of 
            B59 (gray scale and contours). The protostars are marked with star
            symbols, and labeled following
            \citet[][]{2007ApJ...655..364B}. The green and blue contoured
            regions correspond to the regions identified from the C$^{18}$O
            scatter plot (see text and Fig.~\ref{fig:scatter_b59} left panel),
            and the green and blue departures of the $^{13}$CO integrated
            emission. Though adjacent, these two regions have a temperature
            difference of $\sim$2 K, the blue region being warmer. Note that
            while the colder green region hosts a single protostar (B10) and a
            "starless" C$^{18}$O peak, the warmer blue region has three
            embedded protostellar objects (B11, B7 and B8).}}
	\label{fig:pos_scatter_b59}}
\end{figure} 

For C$^{18}$O, a linear fit to all the datapoints provides a
$X_{\mathrm{C}^{18}\mathrm{O}}=1.82\times10^{22}$~cm$^{-2}$K$^{-1}$km$^{-1}$s
(purple dashed line, Fig.~\ref{fig:scatter_b59} left panel), which will be
used to estimate the masses for the bulk of the cloud. However, points above
and below the purple line, with $I_{\rm{C^{18}O}} > 1.5$~K\,km\,s$^{-1}$, fall
into two spatially distinct regions of B59 (Fig.~\ref{fig:pos_scatter_b59},
green and blue contoured regions respectively). These two regions are of
particular interest due to their optical depth and temperature structure. To
better estimate the masses and virial parameters in these two specific
regions, we used a linear fit to the green and blue models, for
$N(\mathrm{H}_{2}) < 8\times10^{22}$~cm$^{-2}$ and $I_{\rm{C^{18}O}} <
3$~K\,km\,s$^{-1}$, shown as blue and green dashed lines on
Fig.~\ref{fig:scatter_b59} left panel.  These correspond to 
$X_{\mathrm{C}^{18}\mathrm{O}} (\mathrm{green} ) = 2.89 \times
10^{22}$~cm$^{-2}$K$^{-1}$km$^{-1}$s and $X_{\mathrm{C}^{18}\mathrm{O}}
(\mathrm{blue} ) = 1.23 \times 10^{22}$~cm$^{-2}$K$^{-1}$km$^{-1}$s.

\begin{figure*}[!t]
	\centering
	{\renewcommand{\baselinestretch}{1.1}
	\includegraphics[angle=270,width=\textwidth]{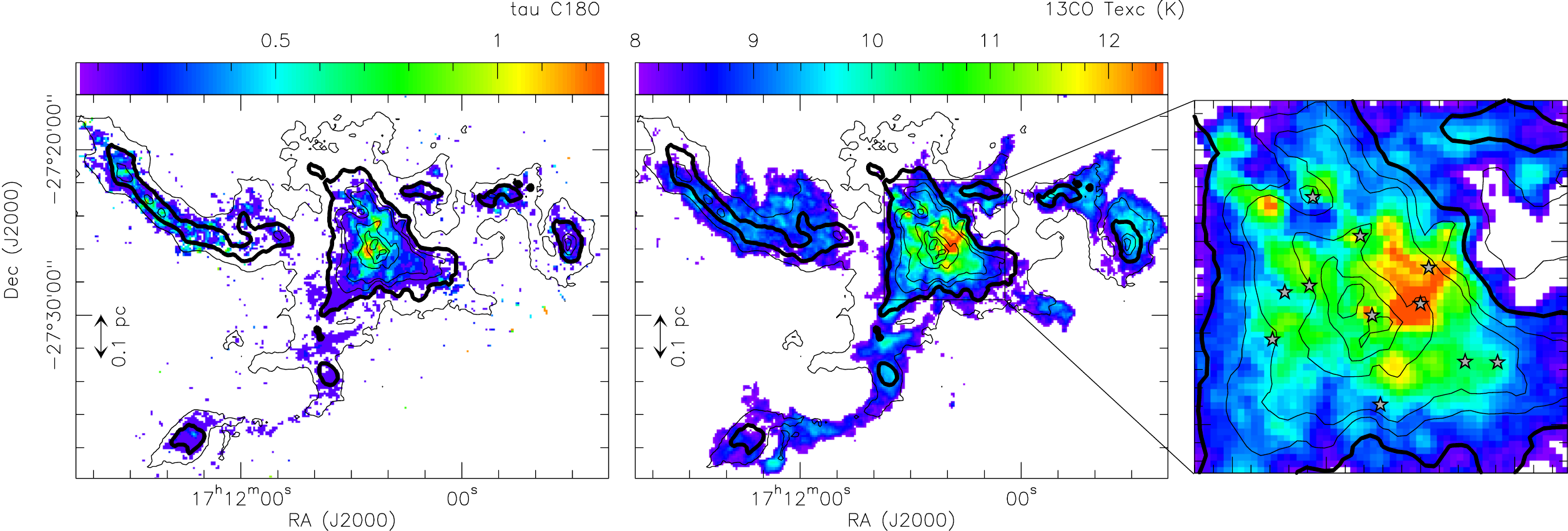}
	\caption[]{\small{\textit{Left}: Colour scale map of the C$^{18}$O
            opacity calculated from the ratio of $^{13}$CO to C$^{18}$O
            emission. The C$^{18}$O optical depth is less than 0.5 in most of
            the cloud and only in the central B59 core it rises to higher
            values, with three local peaks where it reaches values around
            unity. \textit{Centre and right}: $^{13}$CO excitation temperature
            map in colour scale. This shows an increase of temperature towards
            the central B59 (zoomed in on the right panel). In all panels, the
            contours are the dust extinction as in
            Fig.~\ref{fig:b59_ext_sourc}. The darker contour at A$_{v}=10$
            which delineates the regions most likely to be in LTE.}}
	\label{fig:tau_c18o_b59}
	\label{fig:ExcT_13co_b59}}
\end{figure*}

For $^{13}$CO we estimated $X_{^{13}\mathrm{CO}} = 2.85 \times
10^{21}$~cm$^{-2}$K$^{-1}$km$^{-1}$s by constraining the linear fit to pixels
where $N(\mathrm{H}_{2}) < 3\times10^{22}$~cm$^{-2}$ (purple dashed line on
Fig.~\ref{fig:scatter_13cob59} right panel). We use this value to estimate the
mass of the bulk of the cloud, but note that it will only provide a lower
limit in those regions that show significant departures from a linear
relationship due to optical thickening.

\subsection{Optical depth of C$^{18}$O and $^{13}$CO}

To measure the optical depth of the $^{13}$CO and C$^{18}$O and study its
variation across the cloud, we have followed the method in
\citet{1998ApJ...495..871L}. This uses the ratio of the integrated
intensities, and assumes the same excitation temperature for both species.
The method is quite insensitive to the line width assumed, and
therefore we have adopted a velocity width of 1~km\,s$^{-1}$
throughout. 

The map of the ratio of $^{13}$CO to C$^{18}$O integrated intensity (not
shown) has values ranging from 2 to 4 in the denser regions and approaching 9
towards the edges of the cloud.  A linear fit to the pixel-by-pixel comparison
of the two species suggest an abundance ratio of $^{13}$CO to C$^{18}$O of
around 6.4. However this value is significantly affected by the denser regions
where the $^{13}$CO line becomes optically thick.  Therefore, we adopt an
abundance ratio $f$ of $^{13}$CO with respect to C$^{18}$O, of $f=8.4$
\citep[][]{1982ApJ...262..590F}, which is more consistent with the ratio we
find at the edge of the cloud.  This ratio implies C$^{18}$O optical depths
ranging between 0.1 to 0.5 over most of the cloud. However an optical depth
ranging between 0.5 to 1.25 is found towards the central clump
(Fig.~\ref{fig:tau_c18o_b59}), consistent with the expected increase of
opacity from the C$^{18}$O scatter plot.  Therefore, we will consider
C$^{18}$O to be optically thin throughout most of the cloud, but only
marginally so towards the central cores.


\subsection{Excitation temperature structure}
\label{temp}

Since the $^{13}$CO peak emission is optically thick in the bulk of the cloud
($\tau_{13} >$ 1), we estimate the excitation temperature of $^{13}$CO
assuming LTE conditions and optically thick emission. Using the radiative
transfer equations under this assumption \citep[e.g.][]{1986rpa..book.....R},
we have the relation:
\begin{equation}
T_{\mathrm{ex}} (^{13}\mathrm{CO}_{3-2}) = \frac{15.87}{\ln[1+15.87/(T_{max}(^{13}\mathrm{CO})+0.045)]}
\label{eq:Tex}
\end{equation}
where $15.87 = h \nu /k$ for the frequency of $^{13}$CO $J=3\rightarrow2$,
where $h$ is the Planck constant, $\nu$ is the frequency and $k$ is the
Boltzmann constant. Here, $T_{max}$ is the peak main beam temperature of the
line. Where $^{13}$CO is self absorbed, the peak used corresponds to the
absolute maximum of the emission.

\begin{figure*}[!t]
	\centering
	{\renewcommand{\baselinestretch}{1.1}
	\includegraphics[angle=270,width=0.95\textwidth]{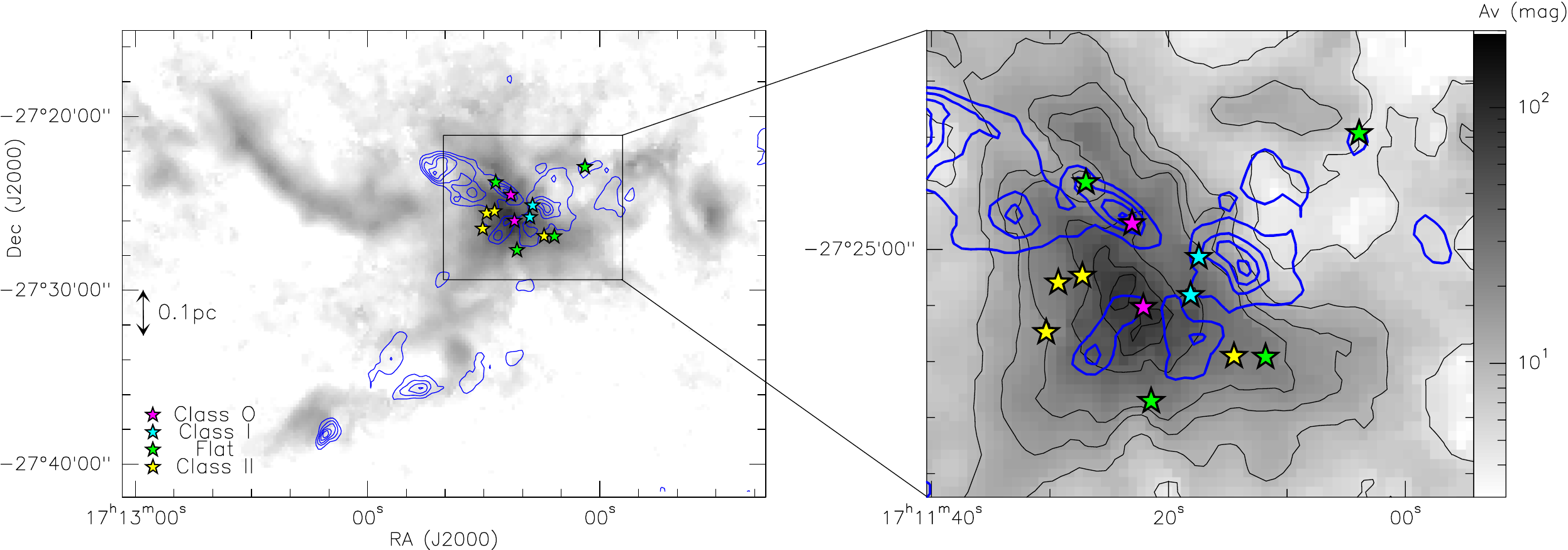}
	\includegraphics[angle=270,width=0.95\textwidth]{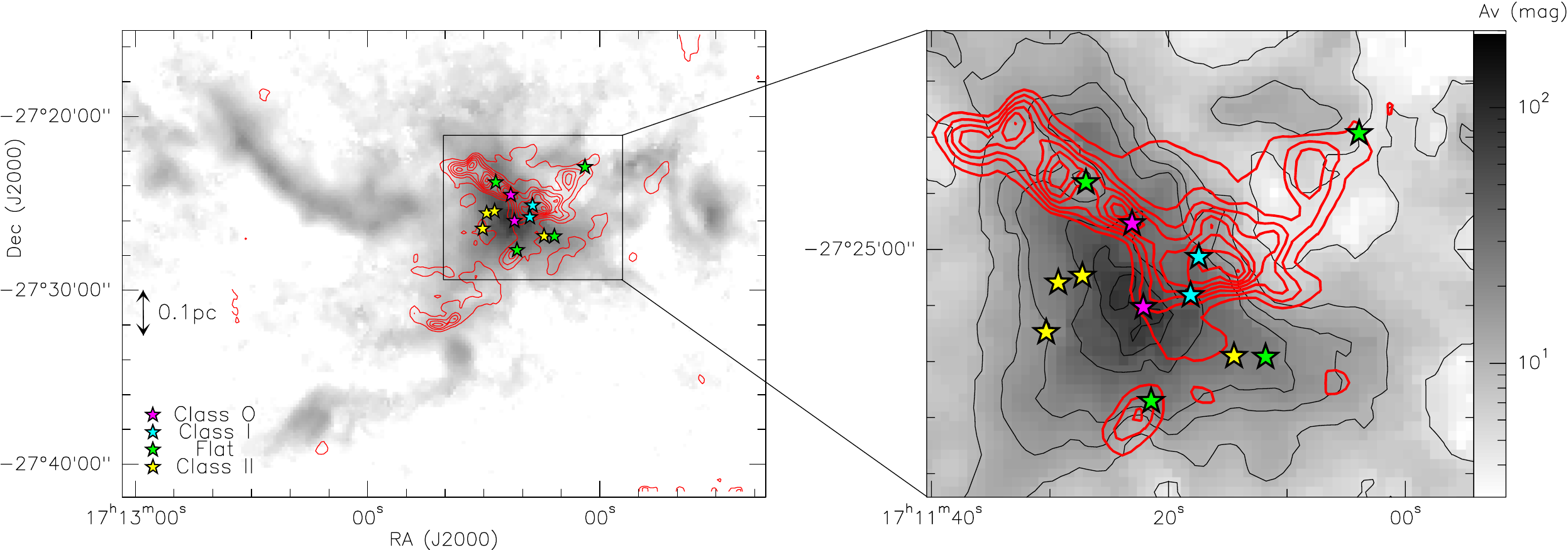}
	\caption[]{\small{$^{12}$CO outflow emission in B59 shown as blue and
            red contours, together with the YSOs positions plotted on the
            extinction map from Fig.~\ref{fig:b59_ext_sourc} in gray scale
            (and black contours on the righthand panels). The lefthand panels
            show the high velocity emission over the entire B59 region, while
            the right panels show a zoom into the central star-forming
            core. The $^{12}$CO emission was integrated from -5 to 2.7~km\,s$^{-1}$ (blue) and from 4.2 to 15~km\,s$^{-1}$ (red). The contours
            are from 3 ~K\,km\,s$^{-1}$ on the left panels and from 4~K\,km\,s$^{-1}$ on the right panels, in steps of 2~K\,km\,s$^{-1}$.}}
	\label{fig:b59_12co_emission}}
\end{figure*} 

The resulting excitation temperature map is shown in
Fig.~\ref{fig:ExcT_13co_b59}, where we can see that most of the cloud is at
$\sim$~9~K, and the star forming core is at 10~-~12~K. Given that the
$^{13}$CO traces some of the outflow emission, in regions with self absorption we will
be measuring the excitation temperature of the warmer outflowing gas. As such,
outflow shocks can be seen in this image as local temperature maxima, even
outside the B59 central region (for instance in the SE U-shaped
ridge). Outside the contoured region shown in Fig.~\ref{fig:ExcT_13co_b59} the
temperatures decrease to as low as 5~K. However, in these diffuse $^{13}$CO
emission regions, not only is it likely that the $^{13}$CO has lower optical
depth, but also the densities are likely lower than the critical densities to
maintain LTE, so the detailed results should be treated with caution.

However we can compare this temperature structure with what is predicted from
the $^{13}$CO scatter plots (right panel of
Fig.~\ref{fig:scatter_13cob59}). Firstly, we can identify three departures
from the linear relation, explained by different gas temperatures (RADEX
models in yellow, green and blue triangles). The first departure, at
$I_{\rm{^{13}CO}} = 2$~K\,km\,s$^{-1}$, corresponds to the NE ridge of B59, and
is consistent with very low temperatures (yellow triangles). The RADEX
predictions indicate temperatures below 7~K, while the LTE estimate
(Fig.~\ref{fig:ExcT_13co_b59}) indicates temperatures closer to 9~K.

The other two departures, one where $I_{\rm{^{13}CO}} \sim$~6~K\,km\,s$^{-1}$ and
the other where $I_{\rm{^{13}CO}}$ has its highest values, at column densities
below $4 \times 10^{22}$~cm$^{-2}$, correspond to the same two regions as
those selected from the C$^{18}$O scatter plot (contoured in
Fig.~\ref{fig:pos_scatter_b59}). These are two adjacent portions in the
central B59 region, each containing one of the two younger sources in the
field (B10 and B11). We find that these two regions show a temperature
difference of $\sim$~2~K in both molecules (green and blue models on
Fig.~\ref{fig:scatter_b59}). Note that the warmer (blue) region also includes
another two more evolved protostars (B7 and B8). This temperature difference
inside the central region is consistent with our LTE $^{13}$CO excitation
temperature estimate shown in Fig.~\ref{fig:ExcT_13co_b59}. The increase of
temperature in the central region is associated with the position of the
protostars, and it is likely due to the combined heating effect from radiative
and outflow feedback from these young sources. However, the importance of
radiative feedback in influencing the temperature is only local (within less
than 10\,000~AU from the protostars), and even though this could change the
fragmentation properties of the gas at these scales, it likely does not
substantially alter the fragmentation at the scale of the B59 star forming
clump (i.e. $\sim$0.3~pc).

\label{H2COdepartures}


\section{Outflows}
\label{b59_outflows}

\subsection{Identification of individual outflows}
\label{sec:out_id}

To identify the population of protostellar outflows in B59, and estimate both
their global and individual physical properties and energetics, we made use of
our $^{12}$CO mapping observations.  These reveal a number of outflows
bursting from the central sources of B59. The lefthand panels of
Fig.~\ref{fig:b59_12co_emission} show the $^{12}$CO blue and red emission from
the larger scale outflows in B59. Here, we see relatively compact outflow
emission bursting to the NE of the central region with spatially coincident
red and blue emission (best seen on the righthand panels of
Fig.~\ref{fig:b59_12co_emission}). This emission falls in the extinction
Cavity B (Fig.~\ref{fig:b59_ext_sourc}) and the outflow appears to trace back
to one of the youngest sources in the field, the Class 0/I source B11.

Another set of larger scale outflow features, including a series of blue
knots, and a red arc can be seen to the SE U-shaped ridge. The zoom-in of the
central region (right panels) show that the driving source of the outflow
responsible for the southern blue knot seems to be the other young Class 0/I
source in the field, source B10.  On the other hand, the red arc seems to be
part of the outflow bursting from an older source (source B9).

This panel also shows that the two Class I sources (B7 and B8, light blue
stars) sit on a region where the $^{12}$CO from the outflows of the two
younger sources (purple stars) start to become confused. Therefore, despite
the fact that some of the $^{12}$CO wing emission in this particular region is
likely to be driven by these two Class I sources, it is hard to disentangle
from the outflows of the other nearby protostars. Finally, the flat spectrum
sources B1/B2 (the western-most green star) also seem to show some signs of
outflow emission. Though not very high velocity nor very strong emission,
these outflow lobes (in particular the red lobe) fall near dust Cavity C.

For many of these outflow features, we cannot clearly identify a counter flow
lobe. This is likely due to confusion of the more compact outflows in the
central region. For this reason, in this paper we will focus on the properties
of the unconfused outflows as well as on the global outflow
properties. Interferometry will be important for disentangling these compact
outflows, identifying their driving sources, and provide a better estimate of
their energetics.

\begin{table*}[!ht]
	\footnotesize
	\centering
	\caption{\small Outflow properties}
		\begin{tabular}{c c c c c c}
		\hline 
		\hline
		 & M  & $p_{\rm{out}}$ 	 & $p_{\rm{out}}^{\rm{corr}}$    & $E_{\rm{out}}$ & $E_{\rm{out}}^{\rm{corr}}$ \\ 
		 & (M$_{\odot}$) & (M$_{\odot}$~km\,s$^{-1}$)  & (M$_{\odot}$~km\,s$^{-1}$) & (M$_{\odot}$~km$^{2}$s$^{-2}$) & (M$_{\odot}$~km$^{2}$s$^{-2}$)\\
		\hline
		Blue & 0.59 & 0.83 & 3.21 & 0.81 & 12.1 \\
		Red	 & 0.63 & 0.55 & 2.13 & 0.40 & 5.97 \\
		\hline
		\end{tabular}
	\label{tab:outflow_prop}
\end{table*}

\begin{table*}[!ht]
	\footnotesize
	\centering
	\caption{\small Momentum flux of individual outflows}
	\begin{tabular}{l c c c c c c c c c c c}
		\hline 
		\hline
		  & Driving & $l$     & $v_{\rm{max}}$ & $i$     & $l^{\rm{ corr}}$ 	 & $v_{\rm{max}}^{\rm{corr}}$ & $t_{d}$    & M          & $p_{\rm{out}}^{\rm{corr}}$    & $E_{\rm{out}}^{\rm{corr}}$ & $F_{out}^{\rm{corr}}$ \\ 
		 & source & (pc)     & (km\,s$^{-1}$) & ($^{\circ}$) & (pc) & (km\,s$^{-1}$)     & ($\times 10^{3}$ yr) & (M$_{\odot}$) & (M$_{\odot}$~km\,s$^{-1}$) & (M$_{\odot}$~km$^{2}$s$^{-2}$) & (M$_{\odot}$~km\,s$^{-1}$yr$^{-1}$) \\
		\hline
		Flow 1 - Blue & B11 & 0.19 & -5.5 & 75 & 0.20 & -21.3 & 9.2  & 0.15 & 0.97  & 4.36 & $1.1 \times 10^{-4}$ \\
		Flow 1 - Red  & B11 & 0.11 & 5.0 & 75 &  0.11 & 19.3  & 5.6  & 0.14 & 0.65  & 2.89 & $1.2 \times 10^{-4}$ \\
		Flow 2 - Red  & B9   & 0.23 & 5.5 & \,57$^{*}$ &  0.27 & 10.1 & 26.1 & 0.22 & 0.36  & 0.46  & $1.3 \times 10^{-5}$ \\	
		Flow 3 - Blue & B10 & 0.45 & -4.0 & 75 & 0.47 & 15.5  & 29.6 & 0.17 & 0.88  & 3.11 & $2.9 \times 10^{-5}$ \\
		\hline
		\end{tabular}
                \flushleft
		{\scriptsize {$^{*}$ 
                    Flow 2 does not show a spatial overlay of the  blue and
                    red wings, indicative of it not being as close to the plane of the sky as flows 1 and 3.
                    We therefore adopt a correction factor of 2.9 for
                    the momentum flux of Flow 2 (i.e. $i\approx57^{\circ}$),
                    based what would be expected from a random distribution of
                    inclination angles
                    \citep[][]{1992A&A...261..274C,1996A&A...311..858B}.
                    If instead an inclination of  75$^{\circ}$ was adopted, the momentum flux of Flow 2 would increase by a factor 5.}}
	\label{tab:momflux_prop}
\end{table*}

\begin{figure*}[!t]
	\centering
	{\renewcommand{\baselinestretch}{1.1}
	\includegraphics[angle=270,width=0.9\textwidth]{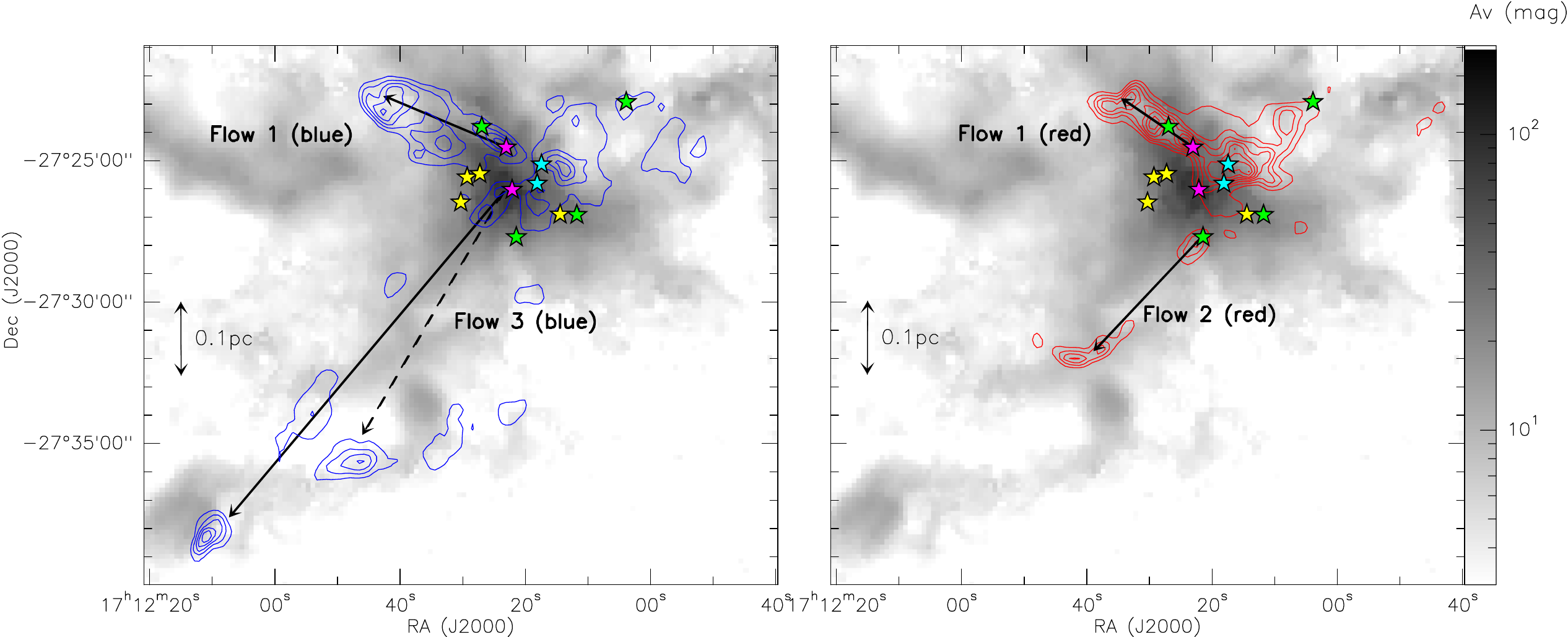}
	\caption[]{\small{Contours of the $^{12}$CO blue-shifted and
            red-shifted emission (left and right panels
            respectively). Contours start at 3~K\,km\,s$^{-1}$ for the blue
            emission and at 4~K\,km\,s$^{-1}$ for the red, with steps of
            2~K\,km\,s$^{-1}$, overlaid on the extinction map of B59 as from
            Fig.~\ref{fig:b59_ext_sourc} in gray scale. The black arrows show
            the different individual outflows used to calculate individual
            outflow properties: Flow 1 (blue and red), Flow 2 (red) and Flow 3
            (blue).}}
	\label{fig:b59_12co_flows}}
\end{figure*} 

\subsection{Outflow inclination angles}
\label{sec:out_ang}

Before estimating the momentum, momentum flux and energy of the outflows, the
observed velocities need to be corrected for the outflow inclination angle $i$
(where $i=0^{\circ}$ is defined to be along the line of sight). {
  Observationally, it is hard to infer the inclination angle of the outflows
  with precision, and this introduces a critical source of uncertainty, as the
  velocities (and momentum) will be affected by a factor of $1/\cos(i)$, the
  energies, by a factor $1/\cos^2(i)$, and the momentum flux, by a factor
  $\sin(i)/\cos^2(i)$.}

{Nevertheless, for a few particular outflows in B59, a careful comparison of
  the blue-shifted and red-shifted emission can shed some light on the overall
  inclination of the flows. In particular, Flow 1
  (Fig~\ref{fig:b59_12co_flows}) powered by B11 shows overlapping blue and
  red emission, consistent with being a single outflow lobe, close to the
  plane of the sky. Because the observed maximum velocities along the line of
  sight for the blue and red wings are similar, this constrains the outflow
  axis to being in the plane of the sky with less than 5$^{\circ}$ uncertainty
  (otherwise the blue and red absolute velocities would differ by more than a
  factor 2). In addition, the projected opening angle of this outflow is
  $\sim$30$^{\circ}$, implying that each of the outflow cone walls are
  inclined by $\sim$15$^{\circ}$ with respect to the plane of the sky, giving
  $i \sim$~75$^{\circ}$. Note that if the outflow axis changes by 5$^{\circ}$,
  the inclination angles would change, but while one would increase to
  80$^{\circ}$, the other would decrease to 70$^{\circ}$. Because the mass,
  extent and velocities of the two lobes are similar, doing so would 
  increase the estimate of the momentum and energy by a factor of
  3. Therefore, $i \sim$~75$^{\circ}$ for Flow 1, remains a rather
  conservative value.}

For Flow 3, the large spatial extent of the collimated blue knot (with a faint
red counterpart at the end), combined with a young driving source (B10) and a
small velocity offset from the ambient cloud, also favours an outflow close to
the plane of the sky. For simplicity, we will adopt the same inclination angle
as for Flow 1.  Considering the preferential alignment of the stronger
outflows close to the plane of the sky, we also adopt $i\sim75^{\circ}$ to
estimate the global outflow properties.

\subsection{Outflow properties}

Estimating the mass of high velocity gas in the outflows, requires correcting
for the optical depth of the $^{12}$CO linewings, which can be done using the
$^{13}$CO data.  
Under LTE and assuming similar excitation temperatures, a
fractional abundance of $^{12}$CO relative to $^{13}$CO of 62
\citep{1993ApJ...408..539L}, and that the $^{13}$CO wing emission is optically
thin \citep[][]{1990ApJ...348..530C}, the optical depth of the $^{12}$CO
($\tau_{12}$) can be determined from the ratio of the integrated wing emission
of the two isotopologues. 

We integrated the emission below 2.7~km\,s$^{-1}$ for the blue wing and above
4.2~km\,s$^{-1}$ for the red.  For each wing, we then apply the correction
factor of $\tau_{12}/(1 - e^{-\tau_{12}})$ for the column density estimate at
each pixel \citep[][]{1990ApJ...348..530C,2010MNRAS.408.1516C} . Assuming a
kinetic temperature of the outflows of 25~K (twice the temperature of the
dense gas), a CO abundance with respect to H$_{2}$ of 10$^{-4}$ and  a
distance of 130~pc, the gas mass can be written as
\begin{equation}
M = 1.78 \times 10^{-6} N_{pix} <\int{T_{mb} dv}>  
\label{eq:mass12}
\end{equation}
where $M$ is the gas mass in unit of M$_{\odot}$ assuming a molecular weight
of 2.33, $N_{pix}$ is the number of pixels in the outflows, and $<\int{T_{mb}
  dv}>$ is the average integrated intensity, after the opacity correction of
the $T_{mb}$ at each pixel. If a higher temperature of 50~K was adopted for
the outflows \citep[as found by, e.g.][]{1999A&A...344..687H}, our mass
estimate would increase by 20\%.

\begin{figure*}[!t]
	\centering
	{\renewcommand{\baselinestretch}{1.1}
	\includegraphics[angle=270,width=0.98\textwidth]{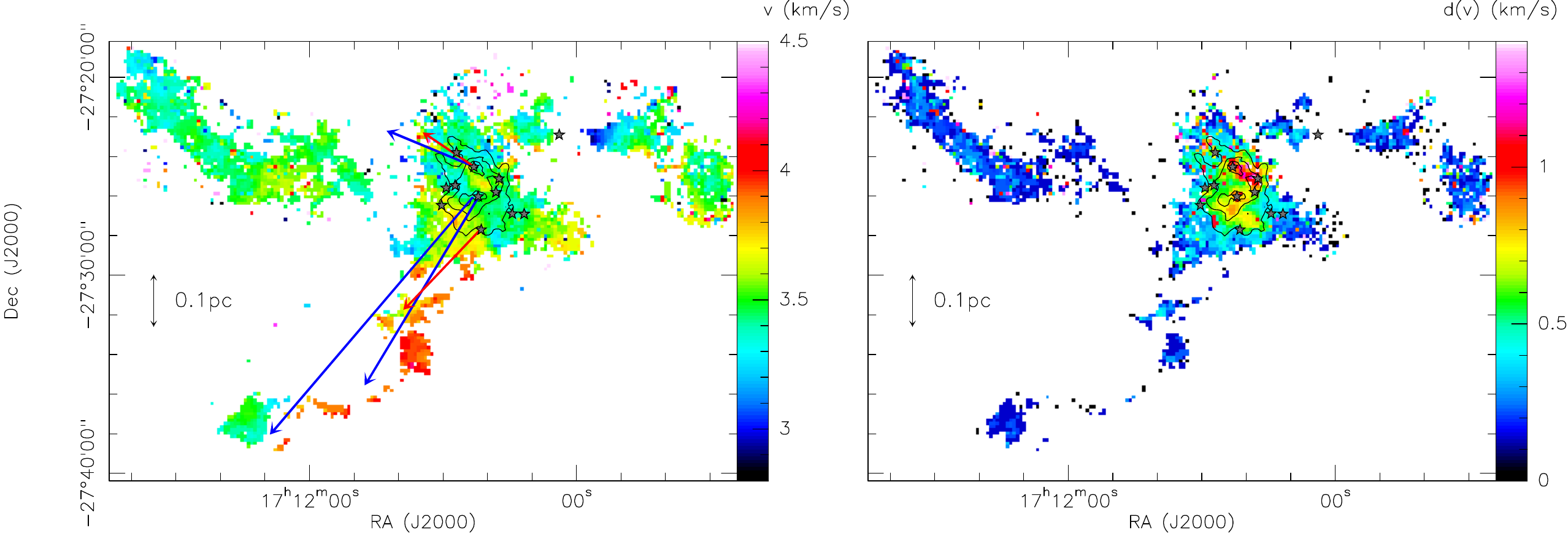}
	\caption[]{\small{C$^{18}$O moment maps: mean velocity on the left
            panel, and velocity dispersion (FWHM) on the right, in colour
            scale overplotted with the integrated intensity in contours, and
            with the sources plotted as grey stars. The outflows identified in
            Fig.~\ref{fig:b59_12co_flows} are also indicated as blue and red
            arrows on the left-hand panel for comparison with the local
            velocity fields.}}
	\label{fig:b59_c18o_momentum}}
\end{figure*} 

To estimate the average momentum, $p_{\rm out}$, and kinetic energy, $E_{\rm
  out}$, of the outflows, as well as the individual outflow momentum flux
,$F_{\rm{out}}$, we use
\begin{equation}
p_{\rm{out}} = \int m(v) |v - v_{0}| dv
\label{eq:moment}
\end{equation}
\begin{equation}
E_{\rm{out}} = \frac{1}{2} \int m(v) (v - v_{0})^{2} dv.
\label{eq:energy}
\end{equation}
\begin{equation}
F_{\rm{out}} = p_{\rm{out}} / t_{d}
\label{eq:momentflux}
\end{equation}
where $v_{0}$ is the velocity of the driving source, $m(v)$ the mass corrected
for the optical depth, and $t_{d}$ is the dynamical time of the outflow,
defined as $t_{d} = l / v_{\rm{max}}$, $l$ being the length of the flow, and
$v_{\rm{max}}$ the maximum velocity of the flow.

The global momentum and energy of the outflows are summarised in
Table~\ref{tab:outflow_prop}, where $M$ is the gas mass contained in each of
the outflow lobes, $p_{\rm{out}}$ is the momentum and $E_{\rm{out}}$ is the
kinetic energy, shown here both without a correction for inclination angle,
and assuming an inclination angle of 75$^{\circ}$
($p_{\rm{out}}^{\rm{corr}}$ and $E_{\rm{out}}^{\rm{corr}}$).

Table~\ref{tab:momflux_prop} shows the individual momentum flux for the four
main flows of B59 (Fig~\ref{fig:b59_12co_flows}). In the table $l$ is the
projected length of the outflow in the plane of the sky, $i$ is the assumed
inclination angle, $v_{\rm{max}}$ is the line-of-sight maximum velocity with
respect to the cloud's ambient velocity of 3.5~km\,s$^{-1}$. The following
columns give the remaining outflow properties, corrected for the inclination
angle. The momentum flux of Flow 1 is higher than the momentum flux of both
Flows 2 and 3 and is consistent with the expected values for Class 0 sources
\citep[i.e. $\sim 10^{-4}$~M$_{\odot}$~km\,s$^{-1}$yr$^{-1}$,
][]{1996A&A...311..858B,2010MNRAS.408.1516C} . Flows 2 and 3 have momentum
fluxes similar to those of a Class I protostar \citep[i.e. $\sim
10^{-5}$~M$_{\odot}$~km\,s$^{-1}$yr$^{-1}$,
][]{1996A&A...311..858B,2010MNRAS.408.1516C}.


\section{Ambient cloud}
\label{ambient}

\subsection{Dynamics}

To study the dynamics of the ambient gas of B59, we use the combined
information from all three CO isotopologues. With the C$^{18}$O emission in
particular, we investigate the velocity { gradients and} changes in the
velocity dispersion across { the region} (Fig.~\ref{fig:b59_c18o_momentum}).
The velocity map of the region (left panel of
Fig.~\ref{fig:b59_c18o_momentum}) shows that the NE ridge is uniform in
velocity, at around 3.5~km\,s$^{-1}$. Not only do we find little variation in
velocity in this ridge, but also the linewidth is very small (with a FWHM as
low as 0.3~km\,s$^{-1}$, right panel of Fig.~\ref{fig:b59_c18o_momentum}).

\begin{figure*}[!ht]
	\centering
	{\renewcommand{\baselinestretch}{1.1}
	\includegraphics[angle=270,width=\textwidth]{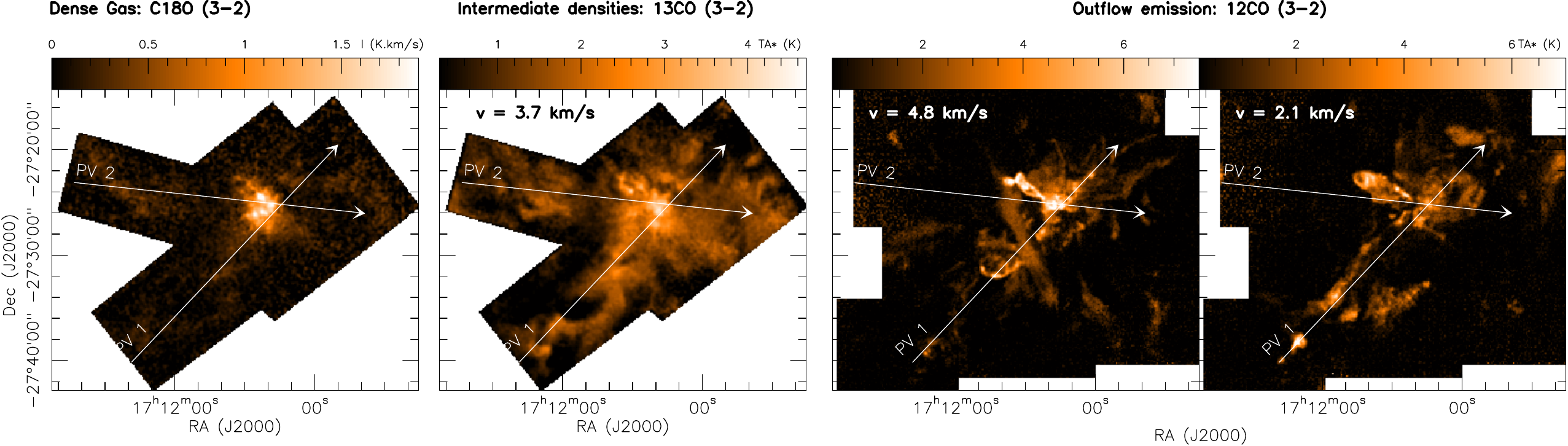}
	\caption[]{\small{Left: Integrated intensity ($\int {\rm
              T}_{\mathrm{A}}^* d\,v$) map of C$^{18}$O showing the
            distribution of the dense gas in B59; Centre-left: $^{13}$CO
            channel map at v~=~3.7~km\,s$^{-1}$ (slightly red-shifted with
            respect to the cloud's ambient velocities of 3.5~km\,s$^{-1}$),
            tracing the intermediate density gas, and showing some of the
            cavities discussed in the text; Centre-right and Right: $^{12}$CO
            channel maps at v~=~4.8~km\,s$^{-1}$ and v~=~2.1~km\,s$^{-1}$
            respectively, illustrating the red-shifted and blue-shifted
            outflows of the region. The position of the two position-velocity
            (P-V) diagrams of Fig.~\ref{fig:b59pv} are shown as white arrows and
            labeled as PV~1 and PV~2 in the four panels.}}
	\label{fig:b59pv_position}}
\end{figure*} 

\begin{figure*}[!t]
	\centering
	{\renewcommand{\baselinestretch}{1.1}
	\includegraphics[angle=270,width=0.8\textwidth]{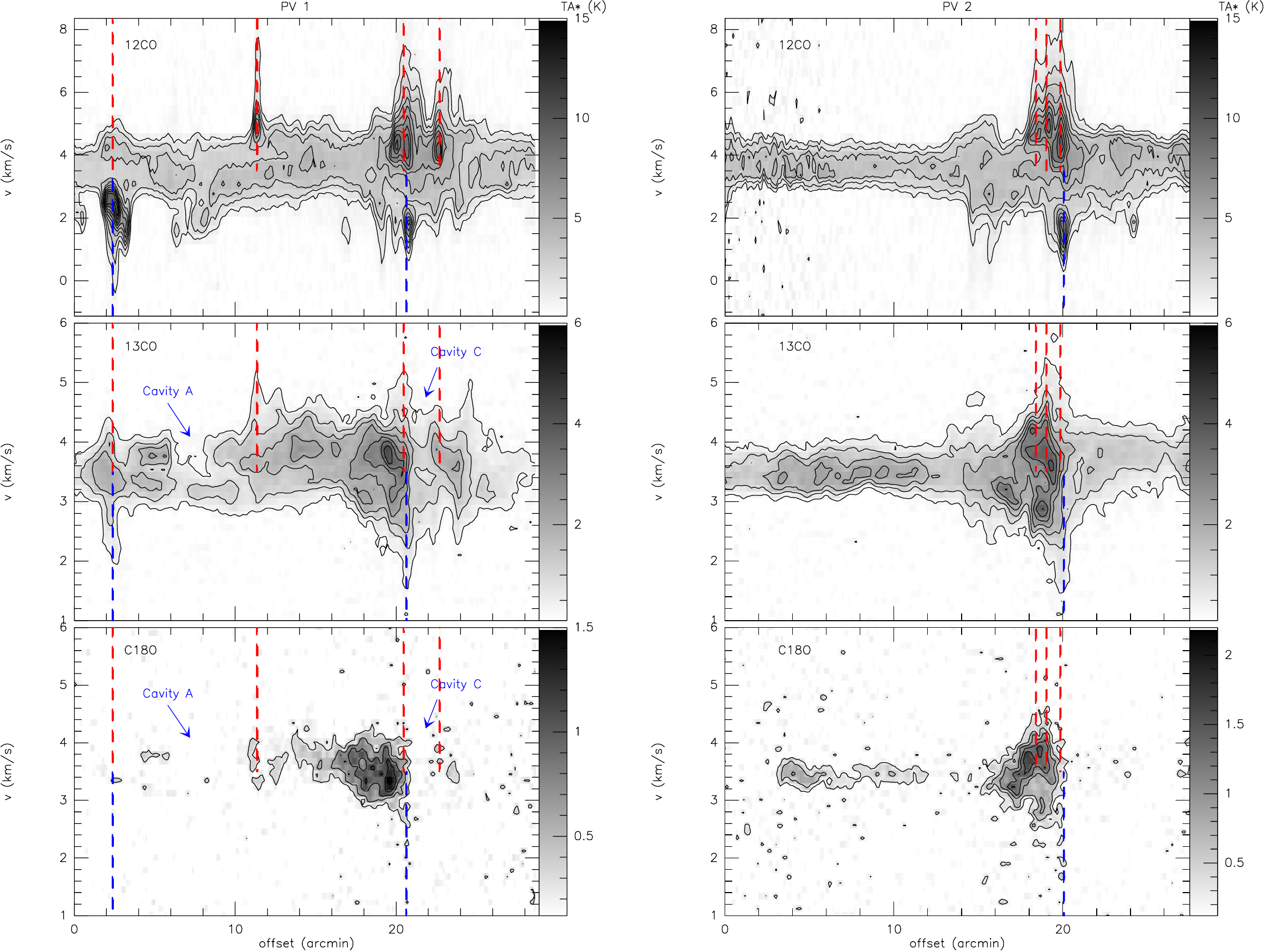}
	\caption[]{\small{P-V diagrams of B59 in $^{12}$CO (top panels),
            $^{13}$CO (centre) and C$^{18}$O (lower panels). Each column is
            the P-V cut as labeled at the top and as shown in
            Fig.~\ref{fig:b59pv_position}. The position of some of the
            outflows are indicated with blue and red dashed lines and cavities
            are also indicated. The contours are from 1~K with steps
            of 1~K for $^{12}$CO, from 0.3~K with steps of 0.6~K for $^{13}$CO
            and from 0.25~K with steps of 0.3~K for C$^{18}$O (in
            T$_{\mathrm{A}}^{*}$).}}
	\label{fig:b59pv}}
\end{figure*}

The U-shaped ridge (Fig.~\ref{fig:b59_ext_sourc}) shows a more complex
velocity structure, with higher velocities than the rest of the cloud
(at around 4~km\,s$^{-1}$). This is somewhat surprising, since this edge
correlates with a blue outflow knot from Flow 3
(Fig.~\ref{fig:b59_12co_emission}). This strengthens the idea that this is an outflow close
to the plane of the sky where 
the redshifted U-shape shows the compressed material behind the outflow, 
although there are no signs of the linewidth broadening in this ridge (right
panel of Fig.~\ref{fig:b59_c18o_momentum}).

The red arc from Flow 2 is also correlated with a C$^{18}$O and extinction
arc, delineating the righthand side of the U-shape. This C$^{18}$O gas has
higher velocities than the centre of B59, consistent with being pushed by the
red-shifted Flow 2 which has a smaller inclination angle than Flows 1 and 3.

In the central clump of B59 there is a velocity gradient of about 1~km\,s$^{-1}$
across 0.1~pc (left panel of Fig.~\ref{fig:b59_c18o_momentum}). {There is
  also a local increase of the linewidth, with the velocity dispersion (FWHM)
  ranging from 0.5~km\,s$^{-1}$ to 1.2~km\,s$^{-1}$}.

This increase of linewidth can also be seen through position-velocity (PV)
diagrams (Fig.~\ref{fig:b59pv_position} and \ref{fig:b59pv}).  The first cut 
(PV 1) is along the axis of Flows 2 and 3. On the left panels of
Fig.~\ref{fig:b59pv} we can see the blue and red outflow knots from Flow 3 at
$\sim$2.1$^{\prime}$ offset as well as the red knot of Flow 2 at
$\sim$11.5$^{\prime}$ offset, in $^{12}$CO and $^{13}$CO.  The velocity
structure of the U-shaped ridge is seen from offsets 3$^{\prime}$ to
14$^{\prime}$ where, while the eastern end is described by velocities
$\sim$3.2~km\,s$^{-1}$, the U-shaped and western end have stronger emission at
$\sim$4~km\,s$^{-1}$ (cf. Fig.~\ref{fig:b59_c18o_momentum}). The $^{13}$CO
traces these two velocities throughout the ridge, but in fact, the U-shape
itself is correlated with the red component (centre-left panel of
Fig.~\ref{fig:b59pv_position}). Cavity A is therefore also seen as a cavity in
the low-velocity redshifted $^{13}$CO emission.  The central region of B59 is
delimited by a sharp edge (at an offset of 20.5$^{\prime}$), where the
emission from C$^{18}$O is broad, and after which the emission drops sharply
down to the noise in Cavity C. This cavity, seen in extinction and with both
$^{13}$CO and C$^{18}$O, is delimited on both sides with high velocity
outflowing gas.

The right column of Fig.~\ref{fig:b59pv} shows the cut through the NE ridge of
B59 (PV 2, from 1$^{\prime}$ to 12$^{\prime}$ offset), confirming the small
linewidths and constant velocity throughout. The emission from the central
portion of B59 is seen at offsets between 14$^{\prime}$ and 20$^{\prime}$ and,
similarly to PV 1, is delimited by a sharp edge to the west (leading to the
same Cavity C as seen in PV 1). The C$^{18}$O emission shows significant local
broadening and the velocities agree with those of $^{13}$CO. In particular,
the red outflow wings correlate with regions where the C$^{18}$O is broader
with stronger red emission. The marked blue outflow wing can be tentatively
seen with C$^{18}$O but it approaches the noise level of { the
  data}. Nevertheless, the lower velocity blue shifted $^{13}$CO emission at an
offset of $\sim$20$^{\prime}$ is also seen in the C$^{18}$O PV diagram.

\subsection{The hierarchical structure}
\label{structure}

To study the hierarchical structure of B59, we have used a dendrogram
extraction technique developed by \citet{2008ApJ...679.1338R}. This method is
useful to understand the different structures that lie within B59. However,
like any other clump-extraction technique, it cannot be blindly used in a
region. In the case of B59, most of the leaves retrieved are not "real" cores
- in the sense of structures which are progenitors of future or currently
forming protostars. In fact, only two leaves in the central B59 correspond to
actual star-forming cores. The remaining structures trace the two halves of
outflow emission, or less dense material which may or may not be bound and
form stars in the future. A detailed description {of the extraction analysis
and results} {which confirm the broken structure of B59} can be found in
Appendix~\ref{hierar}. In summary:
\begin{itemize}
\item{The SE U-shaped ridge is not a single structure but two different
    arch-like structures, consistent with two compressed fronts from two
    different outflows (Flows 2 and 3). This suggests that the gas once in
    Cavity A has been cleared by Flow 3 and compressed, forming
    the eastern side of the U-shaped ridge.}
\item{The Western cores are physically separated from the main B59 structure,
    consistent with the existence of a gas-deprived cavity (Cavity C). The
    {presence of} outflowing gas possibly associated with the B1/B2 system,
    which is seen at the edges of this cavity, suggests that this cavity was
    possibly cleared by the outflow itself earlier in the evolution of these
    sources.}
\item{The NE ridge is, at present, a coherent structure, gravitationally
    bound, uniform in velocity, and with low velocity dispersion. It could be
    on the verge of forming future prestellar cores;}
\item{And finally, the dendrogram analysis finds that the central region does
    not show evidence for further fragmentation into new cores, apart from the
    two cores surrounding the younger protostars in the field. All the other
    structures found in this central region are components of the outflowing
    gas, particularly along the walls of Flow 1.}
\end{itemize}

\section{Discussion: Outflow Interactions}
\label{discussion:out-dense}

\subsection{Outflows and the Cloud}

Our hierarchical study of the region also provides a reliable mass estimates
throughout the different scales of the cloud. These can be used to estimate the
binding energy at different size scales, which can be compared to the kinetic energy 
carried out by the outflows, to estimate their respective impact. The total kinetic
energy of the outflows (Sect.~\ref{b59_outflows}) is
$\sim$18~M$_{\odot}$~km$^{2}$s$^{-2}$, assuming an inclination angle of
75$^{\circ}$. We estimate the binding energy (i.e. the potential energy) of
the dense material using the C$^{18}$O masses. For 30~M$_{\odot}$ and a radius
of 0.30~pc (Appendix~\ref{hierar}), and assuming a density profile $ \propto
r^{-2}$ \citep[similar to that found by][]{2012ApJ...747..149R}, the
gravitational potential energy is 12~M$_{\odot}$~km$^{2}$s$^{-2}$. The
outflows therefore carry an energy comparable to the binding energy of the
dense material. However, the kinetic energy of the outflows may be deposited
outside the cores, and therefore may not be disrupting the dense cores
themselves.

In fact above we have shown that the outflows are correlated with the shapes
and cavities seen in C$^{18}$O and extinction outside the central region,
indicating that the outflows are able to push, carve and shape the material
(e.g. Sect.~\ref{ambient}). For instance the material along the SE U-shaped
ridge is a fossil structure from Flows 2 and 3. Its U-shape, its
velocity structure, and its local temperature maxima are a result of the
impact of outflows on an originally less dense medium.

\subsection{Outflows and the Dense Gas}

Despite the clear impact on larger scales of the outflows in the B59 cloud,
their importance on the dense gas is harder to disentangle. For instance, in
the denser central region of B59, we see temperature peaks likely due to the
combined effect of radiation from the individual protostars, and outflow
shocks from the ensemble of four protostars
(Fig.~\ref{fig:ExcT_13co_b59}). We detect a broadening of the
C$^{18}$O emission towards the central clump, which could be in principle be
associated with a combination of rotation, {shear}, infall and outflow
motions. But an upper limit for the contribution of infall to the linewidth can be
estimated by assuming that the entire B59 central clump is in free fall
collapse. This would result on an infall velocity of
$\sim0.2-0.3$\,km\,s$^{-1}$ (generating a velocity dispersion of that order or
less), insufficient to explain the observed line widths.

\begin{figure*}[!t]
	\centering
	{\renewcommand{\baselinestretch}{1.1}
	\hbox{\includegraphics[angle=270,width=0.38\textwidth]{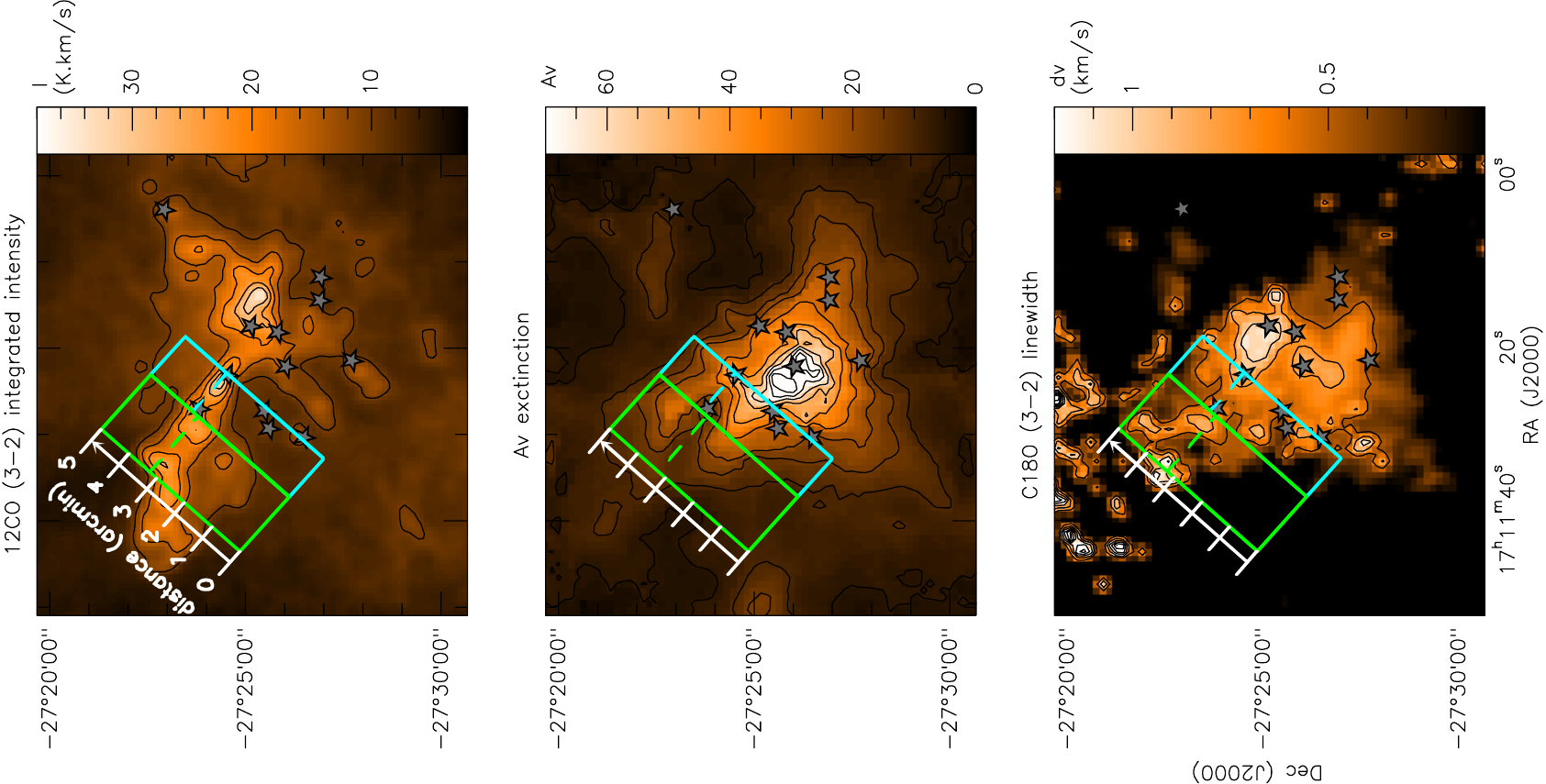}}
	\flushright
	\vspace{-14.4cm}
	\vbox{\includegraphics[width=0.35\textwidth]{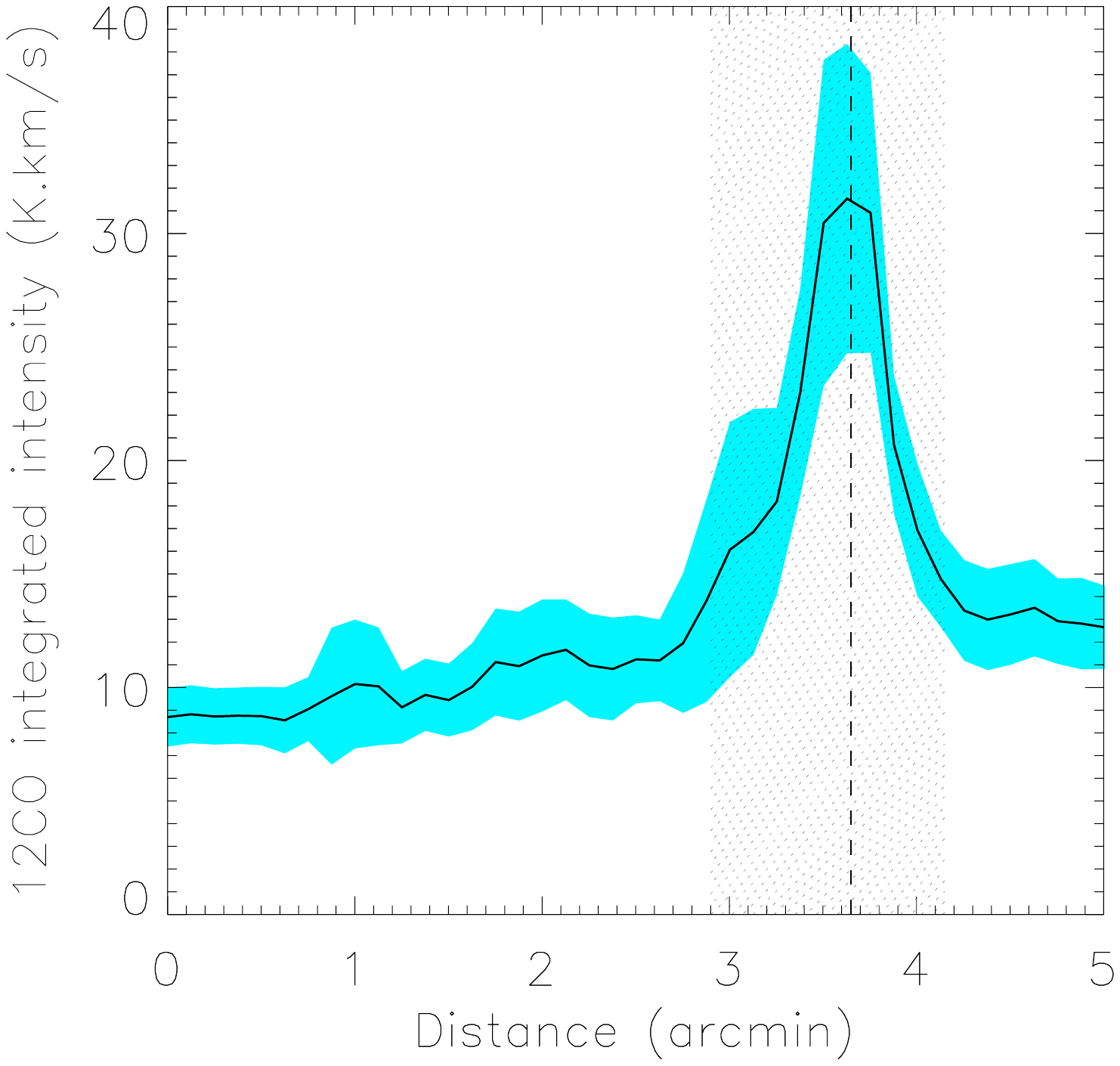}
			\hspace{-1.5cm}
			\includegraphics[width=0.35\textwidth]{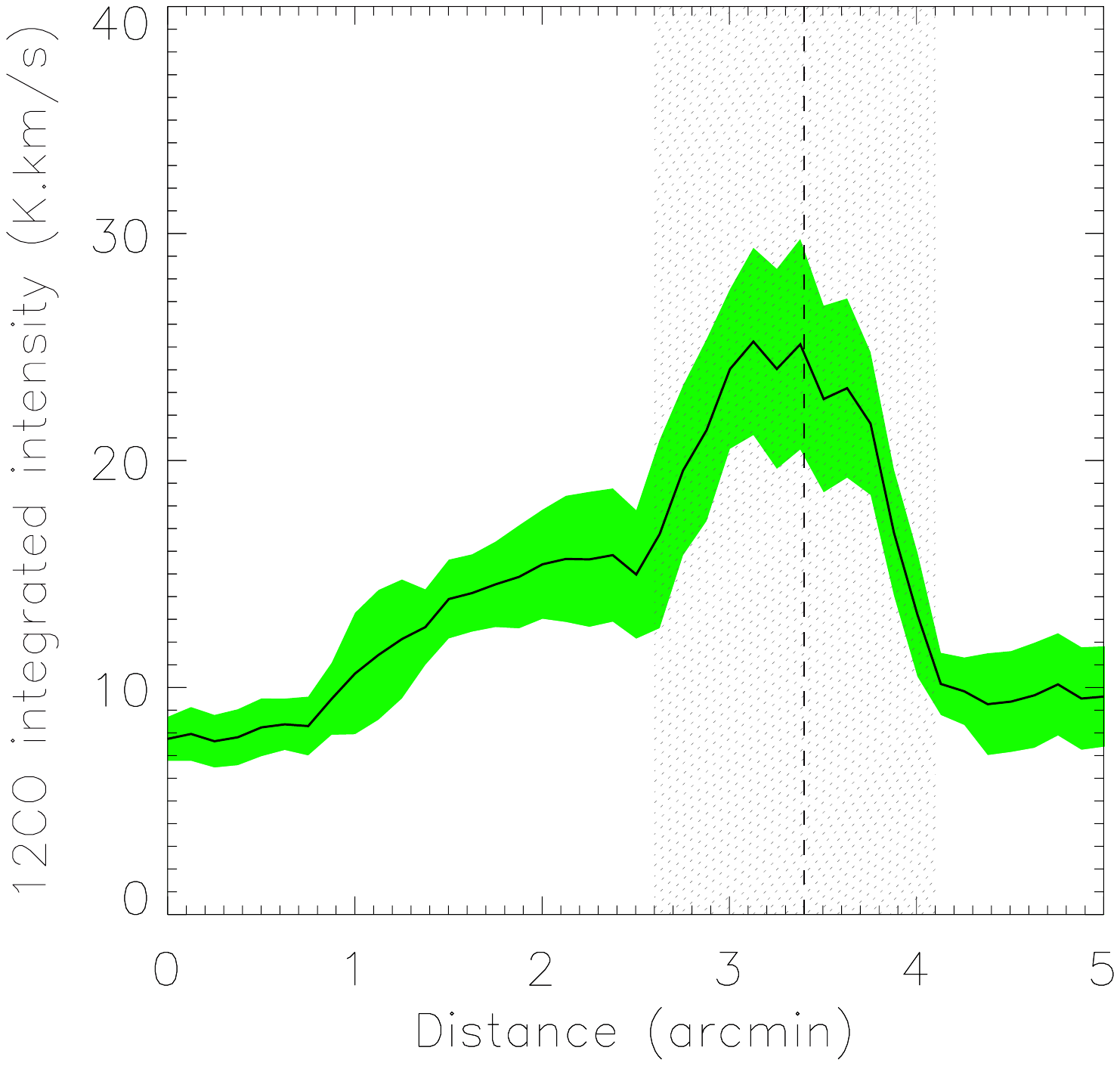}\\
			\vspace{-3.8cm}		
			\includegraphics[width=0.35\textwidth]{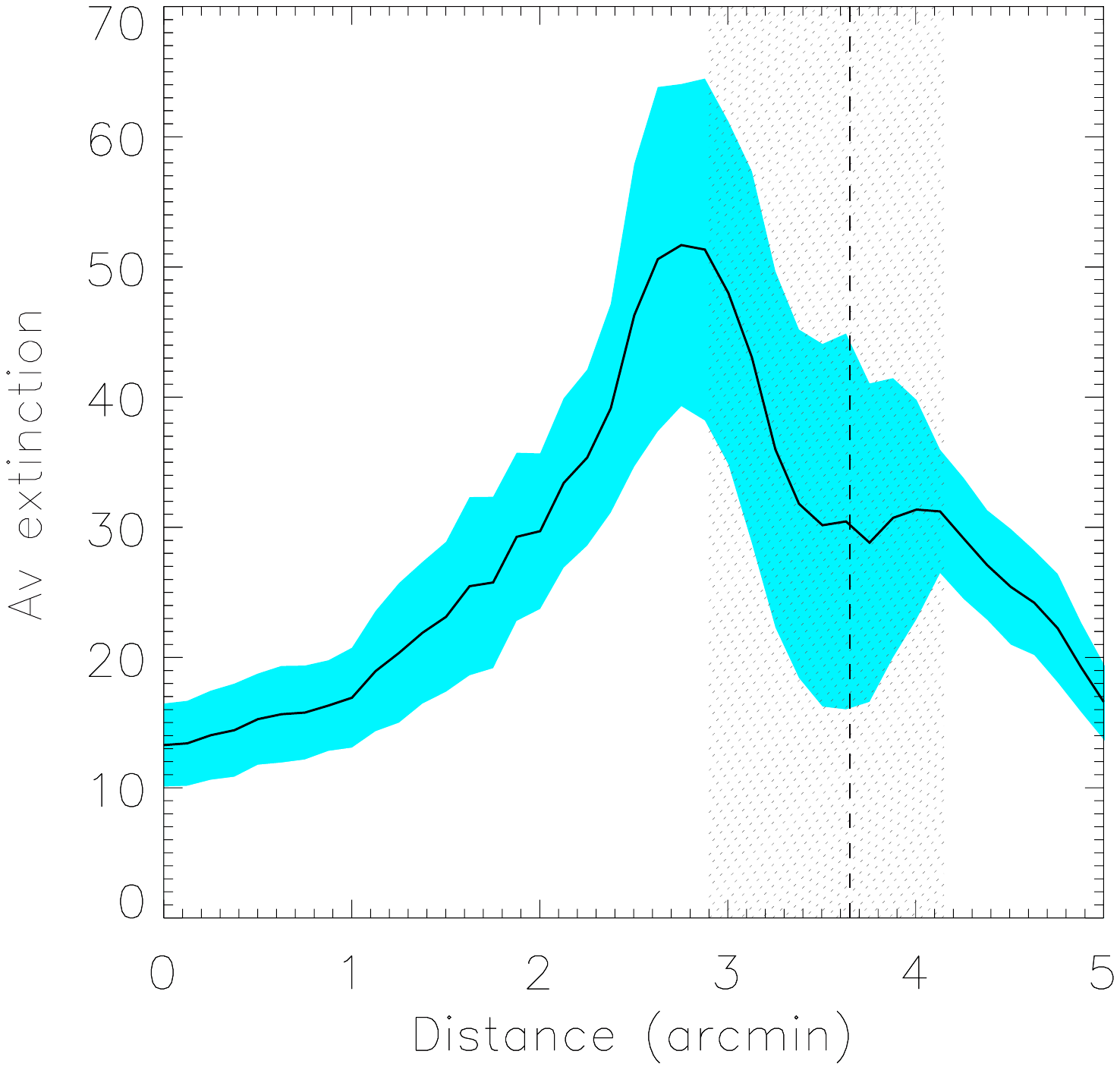}						
			\hspace{-1.5cm}
			\includegraphics[width=0.35\textwidth]{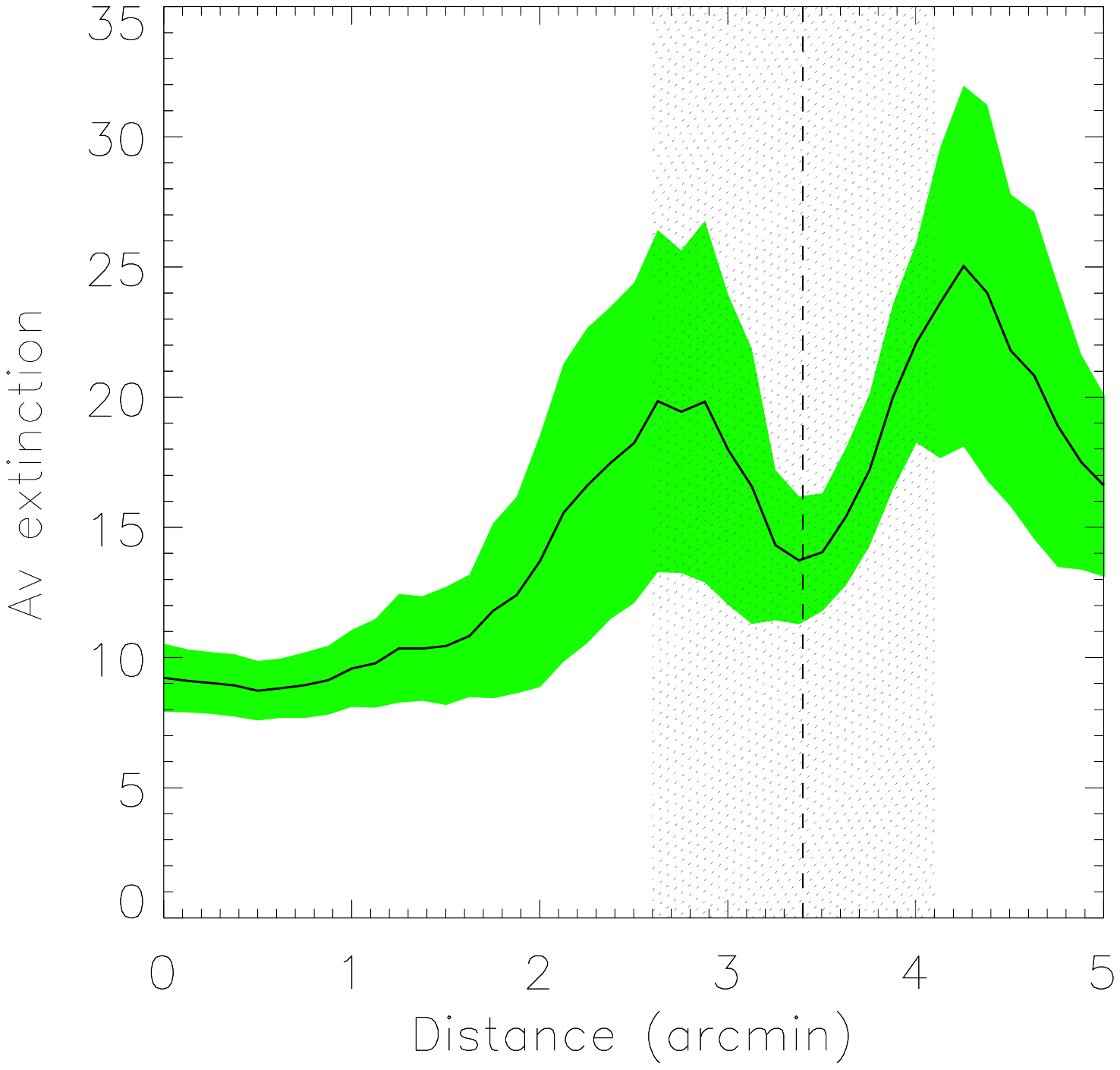}\\	
			\vspace{-3.8cm}	
			\includegraphics[width=0.35\textwidth]{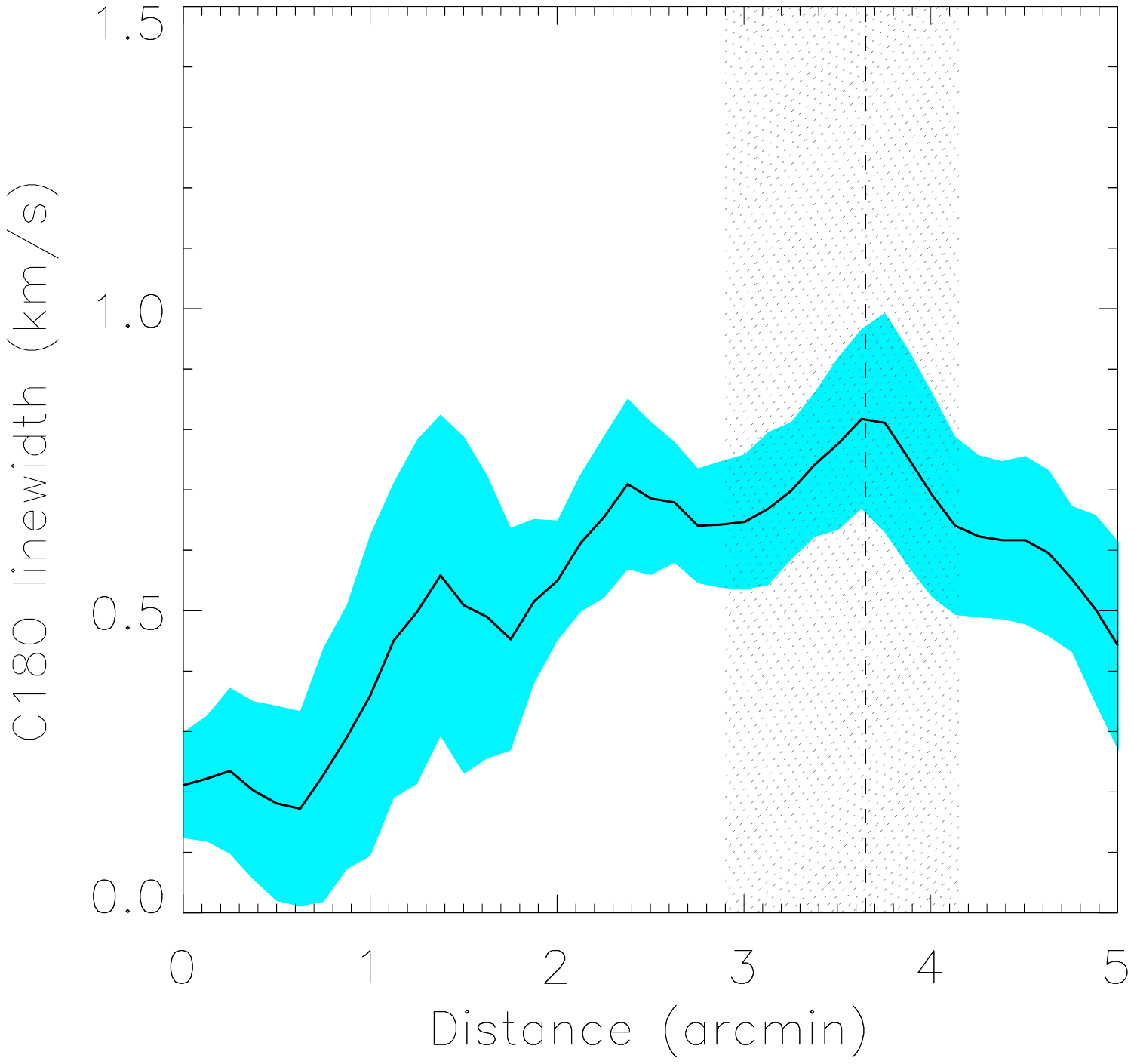}
			\hspace{-1.5cm}
			\includegraphics[width=0.35\textwidth]{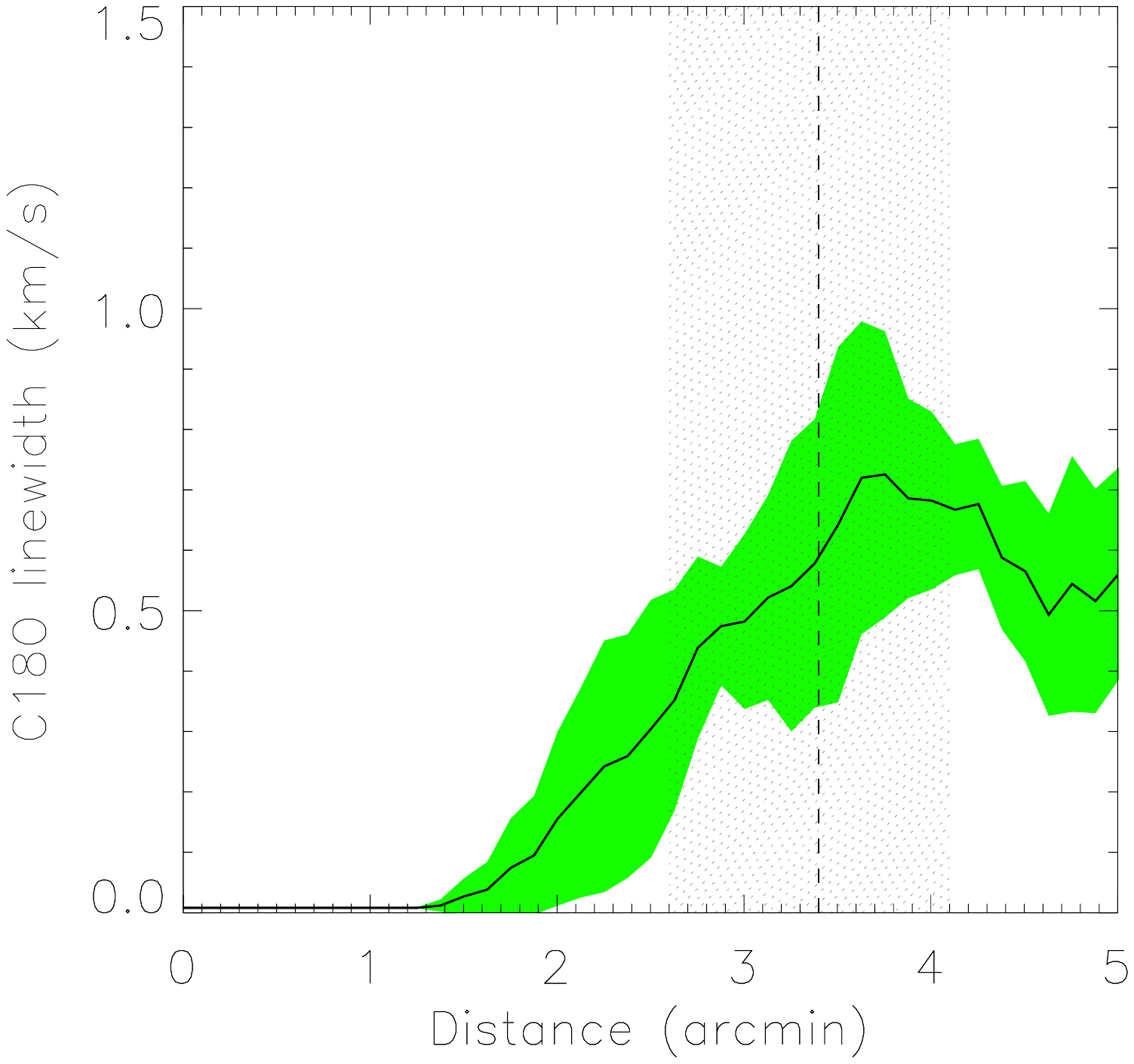}}	
			\vspace{-3cm}
                      \caption{\small{\textit{Left column}: Close up of the
                          central region of B59, where the colour scale and
                          contours are the $^{12}$CO integrated intensity map
                          (top), { the visual extinction map (centre) and the
                            C$^{18}$O linewidth map (bottom)}. The regions
                          used to estimate the average profiles (shown in the
                          centre and right columns) across Flow 1 are shown as
                          blue and green boxes. The dashed lines show the
                          position of the peak of $^{12}$CO integrated
                          intensity as averaged within each
                          box. \textit{Centre and Right columns}: Average
                          profiles of $^{12}$CO integrated intensity (top row
                          panels), { visual extinction (central row panels),
                            and C$^{18}$O linewidth (FWHM, bottom row
                            panels)}. Each column corresponds to a different
                          averaging box, colour-coded as in the left
                          column. The dashed lines { and gray-shadowed areas}
                          mark the peak { and extent} of the $^{12}$CO profile
                          as seen on the first row panels, respectively. The
                          colour-coded shaded areas show the 1-sigma
                          dispersion of each quantity.}}
	\label{fig:outflow_profiles}
	\label{fig:boxes_outflow}}
\end{figure*}

To investigate more closely the possible effect of the outflows on the line width of dense gas,
we have studied the emission in
an area where there is initial evidence for the interaction of the outflow
with the dense gas (in the form of a cavity) and where no significant velocity
gradients are detected. This is the case along Flow 1, which is an unconfused
and well collimated outflow located in the densest part of B59
(Fig.~\ref{fig:b59_12co_emission}). For this test case, we compared the
average profiles of the $^{12}$CO integrated intensity as a measure of the
outflow strength and location; the visual extinction as a measure of the
quantity of material along the line of sight; and the C$^{18}$O linewidth as a
measure of the velocity dispersion.  To do this, we took several slices along
a direction perpendicular to the outflow main axis.


Since the outflow profile changes along its length, we estimated the average
profiles within two separate regions. These are shown as boxes in
Fig.~\ref{fig:boxes_outflow} (left panels) overplotted on the central region
of B59 in the $^{12}$CO integrated intensity (top), { the extinction map
  (middle) and the C$^{18}$O linewidth (bottom)}. These boxes are colour-coded
as the respective profiles, shown in the other panels of the same figure. To
serve as reference, the dashed lines show the position of the peak of
$^{12}$CO integrated intensity.

Closest to the driving source, inside the blue boxed region, the outflow is
collimated and constrained to a $\sim$1'-wide region (top row, central panel
of Fig.~\ref{fig:boxes_outflow}). In the second region, delimited by the green
box, the outflow is broader (top row, right panel). Here, the blue shifted
emission starts forming an arched lobe opening towards the east, 
while the the redshifted outflow ends with a sharp edge
(Fig.~\ref{fig:b59_12co_emission} and \ref{fig:b59_12co_flows}).  This
sharp tip is immediately followed by a linewidth peak, reminiscent of an
impacted outflow knot (Fig.~\ref{fig:boxes_outflow}, first column, {lower}
panel).

In both boxes, the outflow falls in a similar-sized cavity in the extinction
(middle panels). 
This Cavity B (Fig.~\ref{fig:b59_ext_sourc}) has clearly been cleared up by this powerful
  outflow. Figure~\ref{fig:boxes_outflow} (lower row panels) shows the impact of
  the outflow on the dense gas velocity dispersion. 
Even though there is less C$^{18}$O gas in this cavity, there is still enough material to detect an
increase of linewidth along the outflow axis. Immediately
before and after the outflow axis, towards higher column densities, the
linewidths are still relatively high (when compared to the outer parts of
B59), but decrease steadily moving away from the outflowing material. This is
the case for the regions in the green box between 1$^{\prime}$ and
3$^{\prime}$, where the C$^{18}$O linewidth decreases until it reaches our
detection limit. This suggests that the outflows are having a direct impact on
{ both} the shape of the region, and the linewidths of the denser material in
B59.

An upper limit on the contribution from the outflow-generated
turbulence in the total velocity dispersion of the region, can be estimated by assuming
that the initial turbulence of the gas in B59 (before any outflow feedback)
was as low as that currently found outside the central B59 (0.3\,km\,s$^{-1}$
FWHM in the NE ridge, cf. Fig.~\ref{fig:b59_c18o_momentum}, and
Tables~\ref{tab:13co_dendro_prop} and \ref{tab:c18o_dendro_prop}), and that
the increase of linewidth along Flow 1 is solely due to outflow-injected
turbulence. The maximum FWHM detected along Flow 1 is
$\sim$~1.0\,km\,s$^{-1}$, so we estimate that the outflow-generated turbulence
in B59 is locally of the order of 0.9\,km\,s$^{-1}$ FWHM
(i.e. $\sigma_{v}^{out} \sim 0.4$\,km\,s$^{-1}$). This is, however,
most certainly dependent on the power of the outflow and on the age
and mass of the driving source.

Another way to assess the importance of the outflows as turbulence generators
is to compare the turbulent kinetic energy and the outflow energy
\citep[similarly to e.g.][]{2010ApJ...715.1170A}. In the central region, the
kinetic energy is $E_{\rm turb} = 4$~M$_{\odot}$~km$^{2}$s$^{-2}$, while the
outflow energy, estimated in the same area is $E_{\rm out} =
0.62$~M$_{\odot}$~km$^{2}$s$^{-2}$, before applying any correction factor for
outflow inclination. If correcting for a random distribution of angles and
adopting $i\sim$57$^{\circ}$, this becomes $E_{\rm out}^{\rm corr} =
5.2$~M$_{\odot}$~km$^{2}$s$^{-2}$, or even $E_{\rm out}^{\rm corr} = 9.2
$~M$_{\odot}$~km$^{2}$s$^{-2}$ if assuming an inclination angle of
75$^{\circ}$ as in Sect.~\ref{b59_outflows}.  This suggests that the outflows
carry more than enough energy to provide the turbulent support of the central
B59. However, since the central region of B59 shows velocity gradients and is
also likely affected by infall motions, it is not straightforward to conclude
if the observed turbulence is predominantly generated by outflows.

Instead we focus on estimating the fraction of the outflow energy deposited in
the immediate surroundings of Flow 1. Taking the regions in the outflow walls
we estimate from the C$^{18}$O that the total kinetic energy is $E_{\rm turb}
= 0.9$~M$_{\odot}$~km$^{2}$s$^{-2}$. From the respective portion of Flow~1, we
estimate that the outflow energy is $E_{\rm out}^{\rm corr} = 2
$~M$_{\odot}$~km$^{2}$s$^{-2}$. As mentioned in Sect.~\ref{b59_outflows} a
change in the inclination angle adopted for Flow~1 will increase the estimate
of the outflow energy. This together with the fact that the total kinetic
energy around the outflow does include other terms (such as the thermal and
isotropic turbulence), suggests that only a fraction (less than half) of the
energy carried by the outflow is deposited in the immediate surroundings
consistent with, for example, \citet[][]{2006ApJ...646.1059C}. The remaining
energy is deposited outside the star forming core.

We stress that this is estimate is for the specific case of Flow~1. This
efficiency of outflow energy deposition will likely vary depending on the
energy of the outflows and the density and velocity conditions of the regions
with which the outflows interact.


\section{Summary and conclusions}
\label{concl}

B59 is a star forming region with a small and young proto-cluster at its
centre.  In its vicinity the gravitationally bound NE ridge appears likely to
form stars in the future. This ridge has a coherent velocity structure, but
presently shows very little sub-structure.  Another region which may
eventually form stars are the western cores. However the mass reservoir
available in this region is not very large and so if it does become active, it
will likely form no more than a single low mass protostar/protostellar system.
 
All the protostars in B59 are located in a C$^{18}$O twisted filament-like
structure (not seen in the dust) in B59's centre. In this region, there is a
sharp temperature increase of 2K correlated with the position of the
protostars, likely due to radiative heating and outflow
feedback. Nevertheless, the heating is a very localised effect, not likely to
substantially change the properties of the fragmentation of B59 as a whole.

We identify a number of cavities in the B59 region that are either correlated
with outflowing gas, or adjacent to outflow knots. Finally, we investigated
the relation between the outflowing gas and the linewidth of the denser gas
and found a clear correlation with an increase of linewidth of C$^{18}$O along
the outflow axis. \\

Our study of B59 suggests that outflows are responsible for sweeping up and
compressing gas as they make their way out of the cloud. We conclude that the
outflows are interacting closely with the dense material broadening the
linewidths of the material in the densest gas. Outside of the star-forming
core, and as the outflows travel through a lower density material, their
impact is seen as a re-shaping of the gas and dust that surrounds the
protocluster.  Their impact on the dense gas is shown to be an efficient
source of turbulence that provides just enough support for the cloud against
collapse at sub-parsec scales ($\sim$~0.1-0.3~pc). The comparable kinetic and
potential energies are consistent with the idea that the B59 core is
long-lived and slowly collapsing. We, therefore, confirm that outflows are, as
speculated by \citet[][]{2010ApJ...722..971C} and
\citet[][]{2012ApJ...747..149R}, potentially responsible for supporting the
B59 clump against collapse.

A localised line broadening in the dense material is not commonly observed,
since C$^{18}$O does not trace high velocity outflowing gas. In B59, most of
the gas originally had narrow line widths, and the turbulence input from
outflows is more easily detected. In this context, B59 appears in marked
contrast with regions such as Serpens \citep[e.g.][]{2010A&A...519A..27D},
where no clear evidence for the outflows is seen in the C$^{18}$O emission.
Nevertheless, the actual linewidths of C$^{18}$O in Serpens (homogeneously
around $\sim$~1-1.5\,km\,s$^{-1}$ FWHM) are in fact similar to those in the
central core of B59. Serpens being a much more dynamic region resulting from a
collision of clouds \citep[][] {2011A&A...528A..50D} and with a large number
of protostars powering outflows (more than 20 Class 0 and Class I sources in
Serpens against four in B59), it may be that we are simply no longer able to
detect the underlying quiescent gas in Serpens.

We suggest that outflows are an important source of turbulence in the gas
around star forming cores, in agreement with the modeling work from
e.g. \citet[][]{2007ApJ...659.1394M,2007ApJ...662..395N}. However, in regions
where the initial dynamics of the gas implies intrinsically larger linewidths,
the contribution from the outflow-generated turbulence can potentially become
negligible. In fact, the final linewidth will only show a measurable
broadening from the outflow contribution (of at least 30\%), when the injected
turbulence is at least as large as the original initial turbulence of the
gas. Assuming that the measured generated turbulence ($\sigma_{v}^{out}$
$\sim$0.4\,km\,s$^{-1}$) is close to the typical value of Class 0 protostellar
outflows, the outflow broadening will only be detectable in regions where the
original $\sigma_{v}$ of the gas is of the order of 0.4\,km\,s$^{-1}$ (or less). The
relative support provided by such an input will also vary depending on the
mass and the potential energy available to a given protocluster. In
star-forming clumps embedded in more massive molecular clouds (i.e. lying in a
deeper potential well), the gravitational energy will easily exceed such
outflow-induced turbulent support, and the later will not be enough to halt a
larger scale collapse.

We find that the energy carried by the outflows is larger than that needed to
provide the current turbulent support of B59, { and that only a fraction, less
  than half, of the energy of the outflows is efficiently converted into
  turbulent motions of the gas in the immediate surroundings of the
  outflows}. The remaining energy is, therefore, deposited outside the star
forming core. In regions denser than B59 this efficiency may increase, as the
interaction of the outflowing gas and the environment will be more
important. On the other hand, the balance of gravitational and turbulent
energy will also be more important in such denser regions.


\begin{acknowledgements}

  We thank the anonymous referee for his/her comments that helped making the
  paper clearer. We thank J. Pineda and E. Rosolowsky for support with using
  the dendrogram code. ADC is supported by the project PROBeS funded by the
  French National Research Agency (ANR), and also acknowledges the support of
  a grant from the Funda\c{c}\~ao para a Ci\^encia e a Tecnologia (FCT) de
  Portugal, under which part of this work was done. The data reduction and
  analysis was done using GILDAS software (http://www.iram.fr/IRAMFR/GILDAS)
  and Starlink software (http://starlink.jach.hawaii.edu/starlink). The James
  Clerk Maxwell Telescope is operated by the Joint Astronomy Centre (JAC) on
  behalf of the Science and Technology Facilities Council (STFC) of the United
  Kingdom, the Netherlands Organisation for Scientific Research and the
  National Research Council of Canada. This research used the facilities of
  the Canadian Astronomy Data Centre operated by the National Research Council
  of Canada with the support of the Canadian Space Agency.
      
\end{acknowledgements}

\bibliographystyle{aa}	
\bibliography{references}		

\begin{thebibliography}{53}
\expandafter\ifx\csname natexlab\endcsname\relax\def\natexlab#1{#1}\fi

\bibitem[{{Alves} \& {Franco}(2007)}]{2007A&A...470..597A}
{Alves}, F.~O. \& {Franco}, G.~A.~P. 2007, \aap, 470, 597

\bibitem[{{Alves} {et~al.}(2008){Alves}, {Franco}, \&
  {Girart}}]{2008A&A...486L..13A}
{Alves}, F.~O., {Franco}, G.~A.~P., \& {Girart}, J.~M. 2008, \aap, 486, L13

\bibitem[{{Arce} {et~al.}(2010){Arce}, {Borkin}, {Goodman}, {Pineda}, \&
  {Halle}}]{2010ApJ...715.1170A}
{Arce}, H.~G., {Borkin}, M.~A., {Goodman}, A.~A., {Pineda}, J.~E., \& {Halle},
  M.~W. 2010, \apj, 715, 1170

\bibitem[{{Arce} \& {Sargent}(2006)}]{2006ApJ...646.1070A}
{Arce}, H.~G. \& {Sargent}, A.~I. 2006, \apj, 646, 1070

\bibitem[{{Bohlin} {et~al.}(1978){Bohlin}, {Savage}, \&
  {Drake}}]{1978ApJ...224..132B}
{Bohlin}, R.~C., {Savage}, B.~D., \& {Drake}, J.~F. 1978, \apj, 224, 132

\bibitem[{{Bontemps} {et~al.}(1996){Bontemps}, {Andre}, {Terebey}, \&
  {Cabrit}}]{1996A&A...311..858B}
{Bontemps}, S., {Andre}, P., {Terebey}, S., \& {Cabrit}, S. 1996, \aap, 311,
  858

\bibitem[{{Brooke} {et~al.}(2007){Brooke}, {Huard}, {Bourke}, {Boogert},
  {Allen}, {Blake}, {Evans}, {Harvey}, {Koerner}, {Mundy}, {Myers}, {Padgett},
  {Sargent}, {Stapelfeldt}, {van Dishoeck}, {Chapman}, {Cieza}, {Dunham},
  {Lai}, {Porras}, {Spiesman}, {Teuben}, {Young}, {Wahhaj}, \&
  {Lee}}]{2007ApJ...655..364B}
{Brooke}, T.~Y., {Huard}, T.~L., {Bourke}, T.~L., {et~al.} 2007, \apj, 655, 364

\bibitem[{{Buckle} {et~al.}(2009){Buckle}, {Hills}, {Smith}, {Dent}, {Bell},
  {Curtis}, {Dace}, {Gibson}, {Graves}, {Leech}, {Richer}, {Williamson},
  {Withington}, {Yassin}, {Bennett}, {Hastings}, {Laidlaw}, {Lightfoot},
  {Burgess}, {Dewdney}, {Hovey}, {Willis}, {Redman}, {Wooff}, {Berry},
  {Cavanagh}, {Davis}, {Dempsey}, {Friberg}, {Jenness}, {Kackley}, {Rees},
  {Tilanus}, {Walther}, {Zwart}, {Klapwijk}, {Kroug}, \&
  {Zijlstra}}]{2009MNRAS.399.1026B}
{Buckle}, J.~V., {Hills}, R.~E., {Smith}, H., {et~al.} 2009, \mnras, 399, 1026

\bibitem[{{Cabrit} \& {Bertout}(1990)}]{1990ApJ...348..530C}
{Cabrit}, S. \& {Bertout}, C. 1990, \apj, 348, 530

\bibitem[{{Cabrit} \& {Bertout}(1992)}]{1992A&A...261..274C}
{Cabrit}, S. \& {Bertout}, C. 1992, \aap, 261, 274

\bibitem[{{Carroll} {et~al.}(2010){Carroll}, {Frank}, \&
  {Blackman}}]{2010ApJ...722..145C}
{Carroll}, J.~J., {Frank}, A., \& {Blackman}, E.~G. 2010, \apj, 722, 145

\bibitem[{{Carroll} {et~al.}(2009){Carroll}, {Frank}, {Blackman}, {Cunningham},
  \& {Quillen}}]{2009ApJ...695.1376C}
{Carroll}, J.~J., {Frank}, A., {Blackman}, E.~G., {Cunningham}, A.~J., \&
  {Quillen}, A.~C. 2009, \apj, 695, 1376

\bibitem[{{Cavanagh} {et~al.}(2008){Cavanagh}, {Jenness}, {Economou}, \&
  {Currie}}]{2008AN....329..295C}
{Cavanagh}, B., {Jenness}, T., {Economou}, F., \& {Currie}, M.~J. 2008,
  Astronomische Nachrichten, 329, 295

\bibitem[{{Covey} {et~al.}(2010){Covey}, {Lada}, {Rom{\'a}n-Z{\'u}{\~n}iga},
  {Muench}, {Forbrich}, \& {Ascenso}}]{2010ApJ...722..971C}
{Covey}, K.~R., {Lada}, C.~J., {Rom{\'a}n-Z{\'u}{\~n}iga}, C., {et~al.} 2010,
  \apj, 722, 971

\bibitem[{{Cunningham} {et~al.}(2006){Cunningham}, {Frank}, \&
  {Blackman}}]{2006ApJ...646.1059C}
{Cunningham}, A.~J., {Frank}, A., \& {Blackman}, E.~G. 2006, \apj, 646, 1059

\bibitem[{{Curtis} {et~al.}(2010{\natexlab{a}}){Curtis}, {Richer}, \&
  {Buckle}}]{2010MNRAS.401..455C}
{Curtis}, E.~I., {Richer}, J.~S., \& {Buckle}, J.~V. 2010{\natexlab{a}},
  \mnras, 401, 455

\bibitem[{{Curtis} {et~al.}(2010{\natexlab{b}}){Curtis}, {Richer}, {Swift}, \&
  {Williams}}]{2010MNRAS.408.1516C}
{Curtis}, E.~I., {Richer}, J.~S., {Swift}, J.~J., \& {Williams}, J.~P.
  2010{\natexlab{b}}, \mnras, 408, 1516

\bibitem[{{Duarte-Cabral} {et~al.}(2011){Duarte-Cabral}, {Dobbs}, {Peretto}, \&
  {Fuller}}]{2011A&A...528A..50D}
{Duarte-Cabral}, A., {Dobbs}, C.~L., {Peretto}, N., \& {Fuller}, G.~A. 2011,
  \aap, 528, A50+

\bibitem[{{Duarte-Cabral} {et~al.}(2010){Duarte-Cabral}, {Fuller}, {Peretto},
  {Hatchell}, {Ladd}, {Buckle}, {Richer}, \& {Graves}}]{2010A&A...519A..27D}
{Duarte-Cabral}, A., {Fuller}, G.~A., {Peretto}, N., {et~al.} 2010, \aap, 519,
  A27+

\bibitem[{{Forbrich} {et~al.}(2009){Forbrich}, {Lada}, {Muench}, {Alves}, \&
  {Lombardi}}]{2009ApJ...704..292F}
{Forbrich}, J., {Lada}, C.~J., {Muench}, A.~A., {Alves}, J., \& {Lombardi}, M.
  2009, \apj, 704, 292

\bibitem[{{Forbrich} {et~al.}(2010){Forbrich}, {Posselt}, {Covey}, \&
  {Lada}}]{2010ApJ...719..691F}
{Forbrich}, J., {Posselt}, B., {Covey}, K.~R., \& {Lada}, C.~J. 2010, \apj,
  719, 691

\bibitem[{{Frau} {et~al.}(2012){Frau}, {Girart}, \&
  {Beltr{\'a}n}}]{2012A&A...537L...9F}
{Frau}, P., {Girart}, J.~M., \& {Beltr{\'a}n}, M.~T. 2012, \aap, 537, L9

\bibitem[{{Frau} {et~al.}(2010){Frau}, {Girart}, {Beltr{\'a}n}, {Morata},
  {Masqu{\'e}}, {Busquet}, {Alves}, {S{\'a}nchez-Monge}, {Estalella}, \&
  {Franco}}]{2010ApJ...723.1665F}
{Frau}, P., {Girart}, J.~M., {Beltr{\'a}n}, M.~T., {et~al.} 2010, \apj, 723,
  1665

\bibitem[{{Frerking} {et~al.}(1982){Frerking}, {Langer}, \&
  {Wilson}}]{1982ApJ...262..590F}
{Frerking}, M.~A., {Langer}, W.~D., \& {Wilson}, R.~W. 1982, \apj, 262, 590

\bibitem[{{Fuller} \& {Ladd}(2002)}]{2002ApJ...573..699F}
{Fuller}, G.~A. \& {Ladd}, E.~F. 2002, \apj, 573, 699

\bibitem[{{Hatchell} {et~al.}(1999){Hatchell}, {Fuller}, \&
  {Ladd}}]{1999A&A...344..687H}
{Hatchell}, J., {Fuller}, G.~A., \& {Ladd}, E.~F. 1999, \aap, 344, 687

\bibitem[{{Heitsch} {et~al.}(2009){Heitsch}, {Ballesteros-Paredes}, \&
  {Hartmann}}]{2009ApJ...704.1735H}
{Heitsch}, F., {Ballesteros-Paredes}, J., \& {Hartmann}, L. 2009, \apj, 704,
  1735

\bibitem[{{Hennebelle} \& {Chabrier}(2011)}]{2011IAUS..270..159H}
{Hennebelle}, P. \& {Chabrier}, G. 2011, in IAU Symposium, Vol. 270, IAU
  Symposium, ed. {J.~Alves, B.~G.~Elmegreen, J.~M.~Girart, \& V.~Trimble},
  159--168

\bibitem[{{Ladd} {et~al.}(1998){Ladd}, {Fuller}, \&
  {Deane}}]{1998ApJ...495..871L}
{Ladd}, E.~F., {Fuller}, G.~A., \& {Deane}, J.~R. 1998, \apj, 495, 871

\bibitem[{{Langer} \& {Penzias}(1993)}]{1993ApJ...408..539L}
{Langer}, W.~D. \& {Penzias}, A.~A. 1993, \apj, 408, 539

\bibitem[{{Lombardi} {et~al.}(2006){Lombardi}, {Alves}, \&
  {Lada}}]{2006A&A...454..781L}
{Lombardi}, M., {Alves}, J., \& {Lada}, C.~J. 2006, \aap, 454, 781

\bibitem[{{Matzner}(2007)}]{2007ApJ...659.1394M}
{Matzner}, C.~D. 2007, \apj, 659, 1394

\bibitem[{{Matzner} \& {McKee}(2000)}]{2000ApJ...545..364M}
{Matzner}, C.~D. \& {McKee}, C.~F. 2000, \apj, 545, 364

\bibitem[{{Maury} {et~al.}(2009){Maury}, {Andr{\'e}}, \&
  {Li}}]{2009A&A...499..175M}
{Maury}, A.~J., {Andr{\'e}}, P., \& {Li}, Z.-Y. 2009, \aap, 499, 175

\bibitem[{{McKee} \& {Ostriker}(2007)}]{2007ARA&A..45..565M}
{McKee}, C.~F. \& {Ostriker}, E.~C. 2007, \araa, 45, 565

\bibitem[{{Muench} {et~al.}(2007){Muench}, {Lada}, {Rathborne}, {Alves}, \&
  {Lombardi}}]{2007ApJ...671.1820M}
{Muench}, A.~A., {Lada}, C.~J., {Rathborne}, J.~M., {Alves}, J.~F., \&
  {Lombardi}, M. 2007, \apj, 671, 1820

\bibitem[{{Nakamura} \& {Li}(2008)}]{2008ApJ...687..354N}
{Nakamura}, F. \& {Li}, Z. 2008, \apj, 687, 354

\bibitem[{{Nakamura} \& {Li}(2007)}]{2007ApJ...662..395N}
{Nakamura}, F. \& {Li}, Z.-Y. 2007, \apj, 662, 395

\bibitem[{{Nakamura} {et~al.}(2011){Nakamura}, {Sugitani}, {Shimajiri},
  {Tsukagoshi}, {Higuchi}, {Nishiyama}, {Kawabe}, {Takami}, {Karr},
  {Gutermuth}, \& {Wilson}}]{2011ApJ...737...56N}
{Nakamura}, F., {Sugitani}, K., {Shimajiri}, Y., {et~al.} 2011, \apj, 737, 56

\bibitem[{{Onishi} {et~al.}(1999){Onishi}, {Kawamura}, {Abe}, {Yamaguchi},
  {Saito}, {Moriguchi}, {Mizuno}, {Ogawa}, \& {Fukui}}]{1999PASJ...51..871O}
{Onishi}, T., {Kawamura}, A., {Abe}, R., {et~al.} 1999, \pasj, 51, 871

\bibitem[{{Peretto} {et~al.}(2012){Peretto}, {Andr{\'e}}, {K{\"o}nyves},
  {Schneider}, {Arzoumanian}, {Palmeirim}, {Didelon}, {Attard}, {Bernard}, {Di
  Francesco}, {Elia}, {Hennemann}, {Hill}, {Kirk}, {Men'shchikov}, {Motte},
  {Nguyen Luong}, {Roussel}, {Sousbie}, {Testi}, {Ward-Thompson}, {White}, \&
  {Zavagno}}]{2012A&A...541A..63P}
{Peretto}, N., {Andr{\'e}}, P., {K{\"o}nyves}, V., {et~al.} 2012, \aap, 541,
  A63

\bibitem[{{Pineda} {et~al.}(2008){Pineda}, {Caselli}, \&
  {Goodman}}]{2008ApJ...679..481P}
{Pineda}, J.~E., {Caselli}, P., \& {Goodman}, A.~A. 2008, \apj, 679, 481

\bibitem[{{Rathborne} {et~al.}(2008){Rathborne}, {Lada}, {Muench}, {Alves}, \&
  {Lombardi}}]{2008ApJS..174..396R}
{Rathborne}, J.~M., {Lada}, C.~J., {Muench}, A.~A., {Alves}, J.~F., \&
  {Lombardi}, M. 2008, \apjs, 174, 396

\bibitem[{{Riaz} {et~al.}(2009){Riaz}, {Mart{\'{\i}}n}, {Bouy}, \&
  {Tata}}]{2009ApJ...700.1541R}
{Riaz}, B., {Mart{\'{\i}}n}, E.~L., {Bouy}, H., \& {Tata}, R. 2009, \apj, 700,
  1541

\bibitem[{{Rom{\'a}n-Z{\'u}{\~n}iga} {et~al.}(2010){Rom{\'a}n-Z{\'u}{\~n}iga},
  {Alves}, {Lada}, \& {Lombardi}}]{2010ApJ...725.2232R}
{Rom{\'a}n-Z{\'u}{\~n}iga}, C.~G., {Alves}, J.~F., {Lada}, C.~J., \&
  {Lombardi}, M. 2010, \apj, 725, 2232

\bibitem[{{Rom{\'a}n-Z{\'u}{\~n}iga} {et~al.}(2012){Rom{\'a}n-Z{\'u}{\~n}iga},
  {Frau}, {Girart}, \& {Alves}}]{2012ApJ...747..149R}
{Rom{\'a}n-Z{\'u}{\~n}iga}, C.~G., {Frau}, P., {Girart}, J.~M., \& {Alves},
  J.~F. 2012, \apj, 747, 149

\bibitem[{{Rom{\'a}n-Z{\'u}{\~n}iga} {et~al.}(2009){Rom{\'a}n-Z{\'u}{\~n}iga},
  {Lada}, \& {Alves}}]{2009ApJ...704..183R}
{Rom{\'a}n-Z{\'u}{\~n}iga}, C.~G., {Lada}, C.~J., \& {Alves}, J.~F. 2009, \apj,
  704, 183

\bibitem[{{Rosolowsky} {et~al.}(2008){Rosolowsky}, {Pineda}, {Kauffmann}, \&
  {Goodman}}]{2008ApJ...679.1338R}
{Rosolowsky}, E.~W., {Pineda}, J.~E., {Kauffmann}, J., \& {Goodman}, A.~A.
  2008, \apj, 679, 1338

\bibitem[{{Rybicki} \& {Lightman}(1986)}]{1986rpa..book.....R}
{Rybicki}, G.~B. \& {Lightman}, A.~P. 1986, {Radiative Processes in
  Astrophysics}

\bibitem[{{Schneider} {et~al.}(2010){Schneider}, {Csengeri}, {Bontemps},
  {Motte}, {Simon}, {Hennebelle}, {Federrath}, \&
  {Klessen}}]{2010A&A...520A..49S}
{Schneider}, N., {Csengeri}, T., {Bontemps}, S., {et~al.} 2010, \aap, 520, A49+

\bibitem[{{van der Tak} {et~al.}(2007){van der Tak}, {Black}, {Sch{\"o}ier},
  {Jansen}, \& {van Dishoeck}}]{2007A&A...468..627V}
{van der Tak}, F.~F.~S., {Black}, J.~H., {Sch{\"o}ier}, F.~L., {Jansen}, D.~J.,
  \& {van Dishoeck}, E.~F. 2007, \aap, 468, 627

\bibitem[{{Vazquez-Semadeni}(2010)}]{2010arXiv1009.3962V}
{Vazquez-Semadeni}, E. 2010, ArXiv e-prints

\bibitem[{{V{\'a}zquez-Semadeni} {et~al.}(2007){V{\'a}zquez-Semadeni},
  {G{\'o}mez}, {Jappsen}, {Ballesteros-Paredes}, {Gonz{\'a}lez}, \&
  {Klessen}}]{2007ApJ...657..870V}
{V{\'a}zquez-Semadeni}, E., {G{\'o}mez}, G.~C., {Jappsen}, A.~K., {et~al.}
  2007, \apj, 657, 870

\end{thebibliography}

\begin{appendix}

\section{The hierarchical structure of B59}
\label{hierar}

\subsection{Method: Dendrograms}

To study the cloud structure of B59, we used a dendrogram code developed and
applied to star forming regions by \citet{2008ApJ...679.1338R} that traces
the hierarchical structure of the cloud, representing the cloud as a tree-like
structure. This technique begins with the identification of all local maxima
in a datacube (the \textit{leaves} of the tree-like structure), and contours
down the dataset, finding all the underlying structure that connects the
different leaves together (i.e. the \textit{branches}), down to the
\textit{root}. Each connecting point between two structures (leaves, branches
and root) is identified as a \textit{node} in the tree. The code we used to
implement this dendrogram technique is fully described in
\citet{2008ApJ...679.1338R}.  This code also calculates several properties of
the cloud at each contour level, such as the integrated intensity, the size of
the structure, the velocity dispersion, the mean velocity and mean position.

\begin{figure*}[!t]
	\centering
	{\renewcommand{\baselinestretch}{1.1}
	\includegraphics[width=0.5\textwidth]{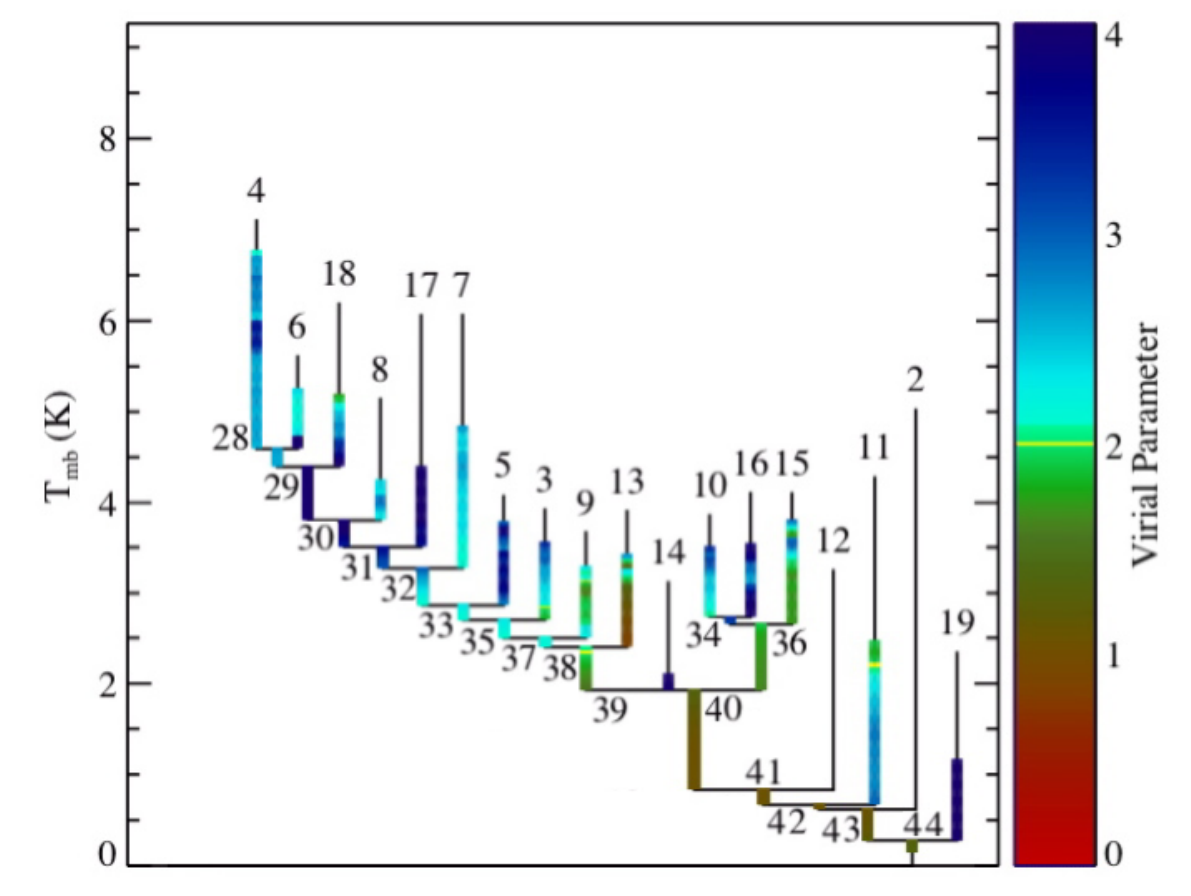}	
	\hfill
	\includegraphics[width=0.37\textwidth]{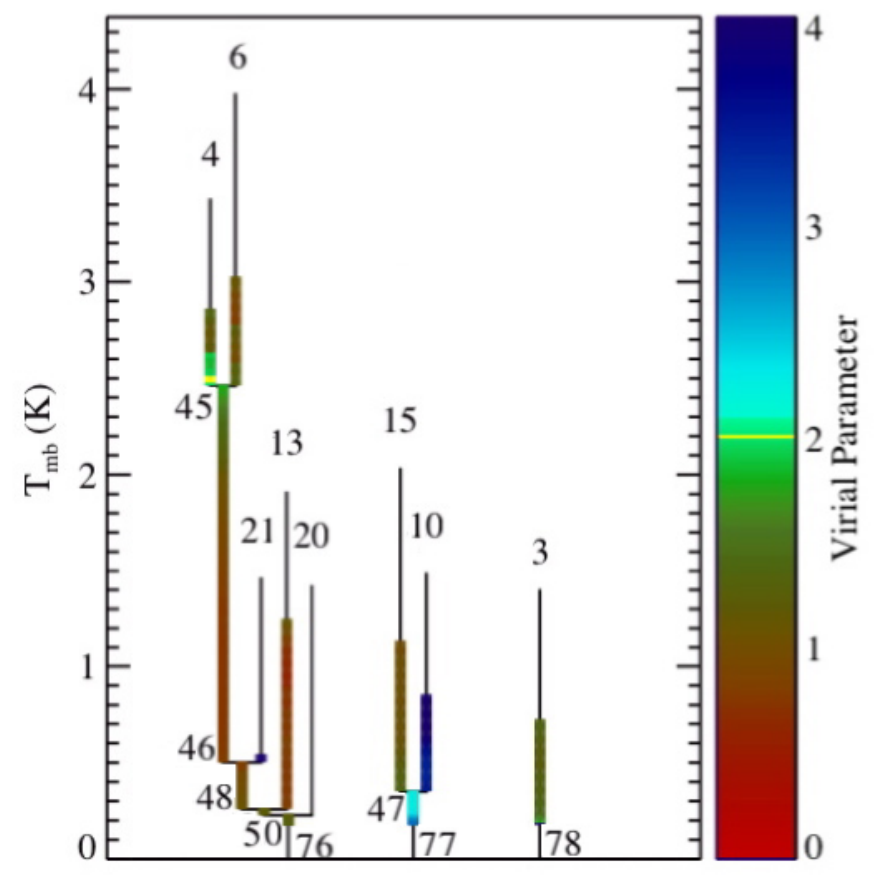}
	\caption{\small{Dendrogram of the $^{13}$CO (left) and C$^{18}$O
            emission (right) in B59, with the virial parameter calculated for
            each contour. The regions where no virial parameter is shown
            (e.g., the tips of the leaves) are due to the lack of sufficient
            quality data to calculate the properties, such as insufficient
            number of pixels. The numbers refer to nodes and leaves as found
            by the code.}}
	\label{fig:dendro_13co}
	\label{fig:dendro_c18o}}
\end{figure*} 

We can calculate the masses within the dendrogram code by taking a linear relation
between the H$_{2}$ column density and the integrated intensity of the
molecular line considered ($X_{\rm{CO}}$), 
following the procedure described in \citet{2008ApJ...679.1338R},
where
\begin{equation}
M = 1.84\times10^{-20} X_{\rm{CO}} L_{\rm{CO}}
\label{eq:mass_dendro}
\end{equation}
where $M$ is the H$_{2}$ gas mass (in M$_{\odot}$) derived from the CO
isotopologue in consideration, $X_{\rm{CO}}$ is the CO-to-H$_{2}$ conversion
factor described in \S~\ref{H2COlinear} (in units of
cm$^{-2}$K$^{-1}$km$^{-1}$s), and $L_{\rm{CO}}$ is the luminosity from the
considered CO emission (in K\,km\,s$^{-1}$pc$^{2}$). This luminosity
$L_{\rm{CO}}$ is calculated as $L=F d^{2}$, where $d$ is the distance to the
cloud and $F$ is the integrated flux of the region, i.e. the sum of all the
emission in the region, calculated as $F=\sum_{i} T_{i}~\delta\theta_{x}
\delta\theta_{y}~\delta v$. The value of $1.84\times10^{-20}$ arises from the
units conversion, assuming a molecular weight of 2.33.

The virial parameter $\alpha$ is estimated for each structure at each contour
level  from the virial balance between the kinetic and gravitational potential
energy, i.e. $\alpha = 2 E_{k}/E_{g}$. Assuming a uniform density profile,
this translates into
\begin{equation}
\alpha = \frac {5 \sigma_{v}^{2} R} {M G} 
\label{eq:virial_dendro}
\end{equation}
where $\sigma_{v}$ is the velocity dispersion and $R$ is the size of the
structure as estimated by the dendrogram code. Assuming a density profile of
$\rho \propto r^{-2}$ would decrease the viral parameter by a factor of
3/5. Without the presence of any additional terms in the virial equation (such
as magnetic field or external pressure), virial equilibrium corresponds to
$\alpha = 1$. However, given the uncertainties associated with these
assumptions, we consider a structure to be gravitationally bound when the
virial parameter is less than 2. The use of such a parameter is best seen as
an internal means of comparison of structures within the cloud rather than an
absolute measure of the individual equilibrium states.

We performed this dendrogram analysis of B59 on the 3D datacubes of the
intermediate density tracers $^{13}$CO and C$^{18}$O. We considered the
minimum height of a leaf to be one-sigma of the r.m.s. noise level, and chose the
bijection method to calculate the cloud properties \citep[similarly
to][]{2008ApJ...679.1338R}.

\subsection{Results}
\label{results_dendro}

\subsubsection{The $^{13}$CO and C$^{18}$O dendrograms}
\label{results_dendro_13co}
\label{results_dendro_c18o}

The dendrograms for $^{13}$CO and C$^{18}$O are shown in
Fig.~\ref{fig:dendro_13co}, left and right panels respectively. This
hierarchical stratification of $^{13}$CO finds the entire structure of B59 to
be interconnected and part of the same cloud, while with C$^{18}$O the cloud
is no longer interconnected (likely due to signal to noise
limitations). Despite the non-interconnectivity, the C$^{18}$O hierarchical
stratification is consistent with that found with $^{13}$CO though with less
branching.

\begin{table}[!t]
	\footnotesize
	\centering
	\caption{\small Properties of the $^{13}$CO dendrogram leaves}
		\begin{tabular}{c | c c c c c c}
		\hline 
		\hline
		Region & Leaf \# & $R$    & $v_{0}$ 		& $\sigma_{v}$ & $M$ & $\alpha$ \\  
			   &		  & (pc) & (km\,s$^{-1}$) 	& (km\,s$^{-1}$)  &(M$_{\odot}$) &  \\  
		\hline
		\hline
		Central 	& 4	& 	0.03 & 3.9 & 0.19 & 0.63 & 2.3 \\
		clump 	& 6	& 	0.02 & 3.7 & 0.13 & 0.10 & 3.6 \\
				& 18	& 	0.02 & 2.9 & 0.13 & 0.10 & 3.7 \\
				& 8	&	0.01 & 3.9 & 0.07 & 0.04 & 2.0 \\
				& 17	&	0.04 & 3.2 & 0.15 & 0.31 & 3.2 \\
				& 7	&	0.03 & 3.7 & 0.11 & 0.19 & 1.9 \\
		\hline
		U-shaped	& 5	&	0.06 & 3.9 & 0.14 & 0.57 & 2.5 \\
		ridge		& 3	&	0.07 & 4.0 & 0.10 & 0.47 & 1.6 \\
		\hline
		NE ridge 	& 13	&	0.14 & 3.5 & 0.13 & 3.70 & 0.8 \\
		\hline	
		Western	& 10	&	0.05 & 3.4 & 0.14 & 0.63 & 1.9 \\
		cores	& 16	&	0.03 & 3.2 & 0.15 & 0.20 & 3.9 \\
				& 15	&	0.06 & 3.4 & 0.15 & 1.15 & 1.5 \\
		\hline
		SW lobe	& 9	&	0.04 & 3.6 & 0.11 & 0.28 & 1.9 \\
		\hline
		SW lobe	& 14	&	0.03 & 3.4 & 0.08 & 0.06 & 3.6 \\
		\hline	
		NW lobe	& 11	&	0.04 & 3.5 & 0.11 & 0.21 & 2.6 \\
		\hline
		NW lobe	& 19 & 	0.04 & 2.2 & 0.20 & 0.13 & 13.6 \\
		\hline
		Total	Centre & (32) & 	0.11 &   -  & 0.33 & 7.87 & 1.9 \\
		\hline
		Total		& (44) & 	0.44 &   -  & 0.41 & 73.3 & 1.1 \\
		\hline
		\end{tabular}
	\label{tab:13co_dendro_prop}
\end{table}

\begin{table}[!t]
	\footnotesize
	\centering
	\caption{\small Properties of the C$^{18}$O dendrogram leaves}
		\begin{tabular}{c | c c c c c c}
		\hline 
		\hline
		Region & Leaf \# & R    & V$_{0}$ 		& $\sigma_{v}$ & $M$ & $\alpha$ \\  
			   &		  & (pc) & (km\,s$^{-1}$) 	& (km\,s$^{-1}$)  &(M$_{\odot}$) &  \\  
		\hline
		\hline
		Central		& 4 & 0.02 & 3.4 & 0.14 & 0.19 & 1.6 \\
		clump		& 6 & 0.03 & 3.5 & 0.21 & 1.44 & 1.1 \\
		\hline
		U-shaped		& 3 & 0.06 & 3.4 & 0.13 & 0.91 & 1.5 \\
		ridge			&   &   &   &  &  &  \\
		\hline
		NE ridge		& 13 & 0.15 & 3.5 & 0.15 & 5.82 & 0.7 \\
		\hline
		Western  		& 15 & 0.07 & 3.4 & 0.15 & 1.51 & 1.2 \\ 
		cores		& 10 & 0.07 & 3.3 & 0.21 & 1.15 & 2.9 \\
		\hline
		West centre 	& 21 & 0.05 & 3.5 & 0.17 & 0.52 & 3.3 \\
		\hline
		Total Centre	& (45)  & 0.10 & - & 0.29 & 16.1 & 0.7 \\ 
		\hline
		Total 		&   & 0.30 & - & 0.31 & 29.6 & 1.0 \\ 
		\hline
		\end{tabular}
	\label{tab:c18o_dendro_prop}
\end{table}

The properties of the dendrogram leaves are summarised in
Tables~\ref{tab:13co_dendro_prop} and ~\ref{tab:c18o_dendro_prop}. The first
column indicates the region to which each leaf belongs and the second
column shows the dendrogram ID number (cf. Fig.~\ref{fig:dendro_13co},
\ref{fig:pos_dendro_13co} and \ref{fig:pos_dendro_c18o}). The following
columns show the properties derived at the lowest contour of each detected
leaf: $R$ is the radius; $v_{0}$ is the mean velocity, weighted by the
intensity (with an uncertainty of 0.1~km\,s$^{-1}$); $\sigma_{v}$ is the
velocity dispersion; $M$ is the leaf mass (eq.~\ref{eq:mass_dendro}); and
$\alpha$ is the virial parameter (eq.~\ref{eq:virial_dendro}). Bear in mind
that the virial parameters shown in this table are estimated at the bottom of
each leaf, but they evolve throughout the height of the leaves, generally
approaching subvirial values (Fig.~\ref{fig:dendro_13co}). In addition, the
row before-last shows the properties of the branch that covers the entire
central region, and the last row shows the total mass, equivalent size and
virial parameter of the entire B59. For $^{13}$CO, the properties of the
entire B59 were measured at the root of the dendrogram (root 44,
Fig.~\ref{fig:dendro_13co} left panel), while for C$^{18}$O this was estimated
by combining the information of the three disconnected roots
(Fig.~\ref{fig:dendro_13co} right panel, roots 50, 47 and leaf 3).

These estimates do not account for the optical depth, which will particularly
affect the central region of B59. Figure~\ref{fig:scatter_b59} quantifies the
possible effect of optical depth for C$^{18}$O as deviations at higher column
densities above the green dashed line. This would account for a difference of
$\sim15\%$ on average in the column density in this central region, which
translates into an equivalent increase of the mass estimate (and decrease of
$\alpha$).

\subsubsection{The leaves of B59}

\begin{figure*}[!ht]
	\centering
	{\renewcommand{\baselinestretch}{1.1}
	\includegraphics[angle=270,width=0.94\textwidth]{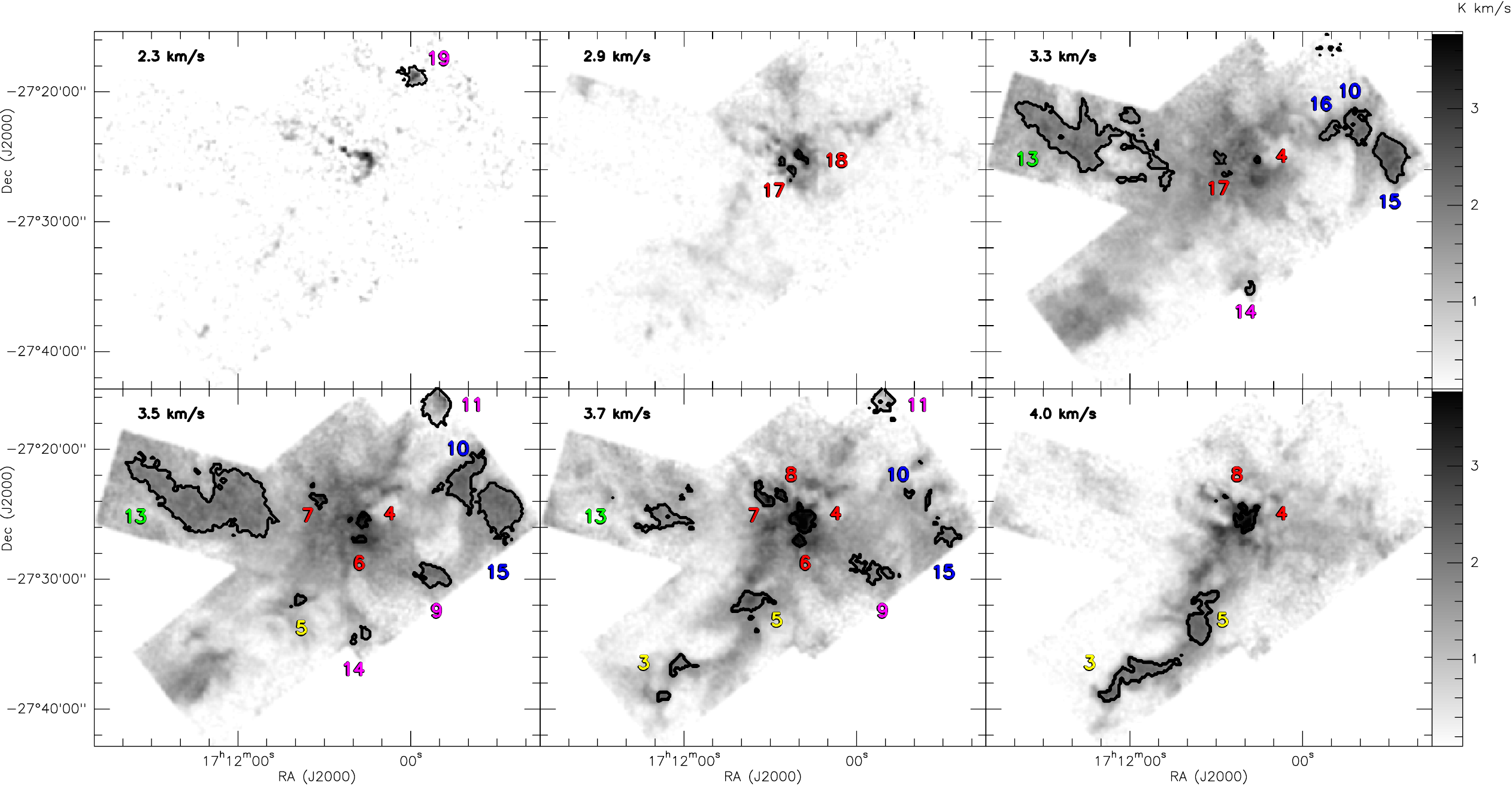}
	\caption[]{\small{Six channel maps of the $^{13}$CO emission,
            T$_{\mathrm{A}}^{*}$, whose velocities are shown at the
            top-left corner of each panel. The plotted grayscale is that used
            for all panels with the exception of the first (top-left panel),
            where it is from 0.1 to 1.5~K\,km\,s$^{-1}$. The black contours show
            the positions of the leaves from Fig.~\ref{fig:dendro_13co}, whose
            numbering is colour-coded to visually separate the distinct areas:
            central region in red, NE ridge in green, SE U-shaped ridge in
            yellow, western cores in blue, and loose knots in purple.}}
	\label{fig:pos_dendro_13co}}
\end{figure*} 

\begin{figure*}[!t]
	\centering
	{\renewcommand{\baselinestretch}{1.1}
	\vspace{-0.2cm}
	\includegraphics[angle=270,width=0.66\textwidth]{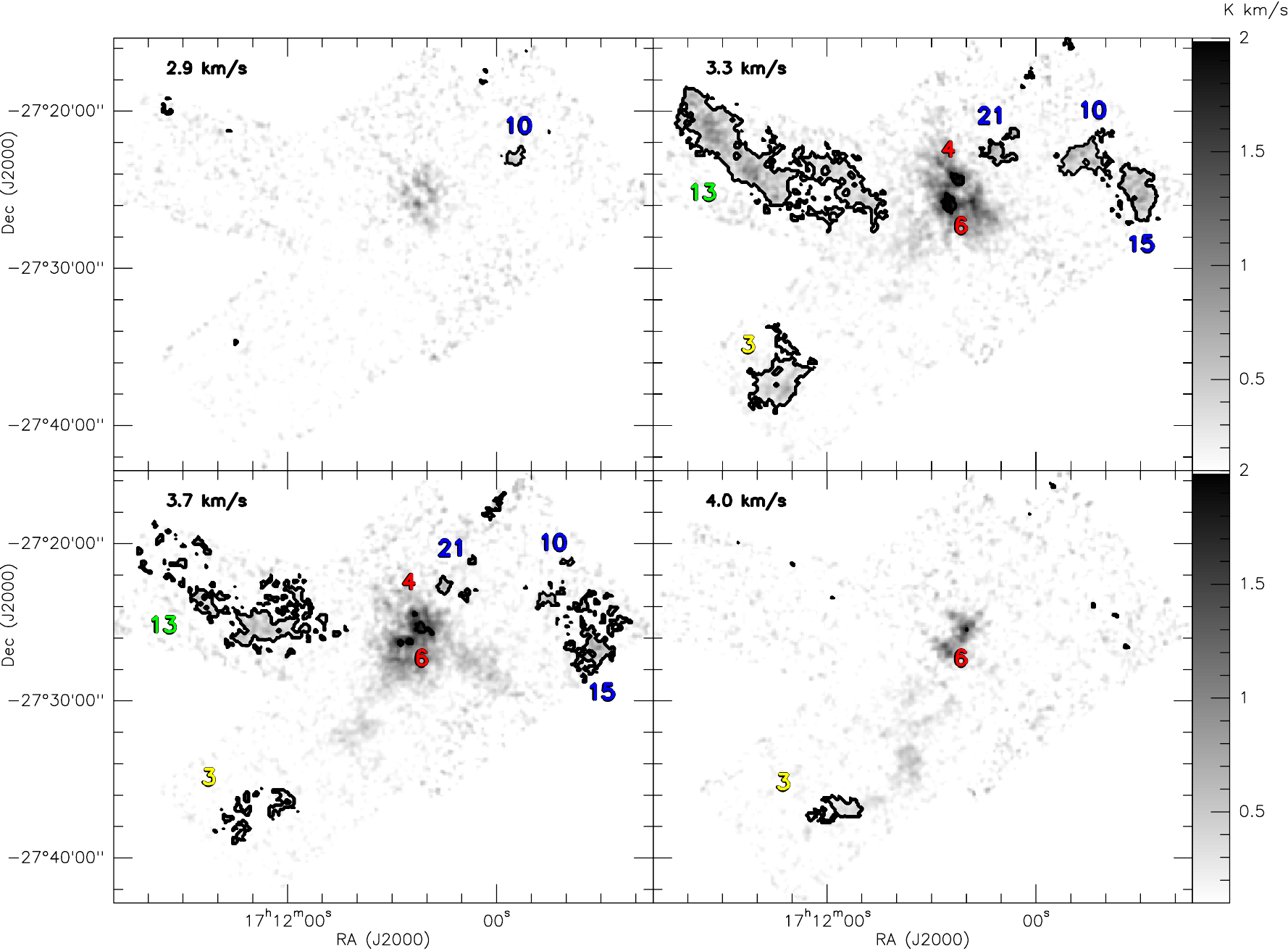}
	\caption[]{\small{Four channel maps of the C$^{18}$O emission in
            T$_{\mathrm{A}}^{*}$ (in gray scale), whose velocities are shown
            at the top-left corner of each panel. The black contours show the
            positions of the leaves from Fig.~\ref{fig:dendro_c18o}, and are
            numbered accordingly. The colour code is the same as for
            Fig.~\ref{fig:pos_dendro_13co}.}}
	\label{fig:pos_dendro_c18o}}
\end{figure*}

The densest gas within the central clump of B59, is solely described by two
leafs in both molecular lines, leaves 4 and 6, both gravitationally bound, and
both at the systemic velocity of the cloud (in C$^{18}$O). These correspond to
the two cores surrounding the youngest protostars of the region: B7, B8 and
B11 within leaf 4, and B10 within leaf 6. Apart from these, we have four
leaves with $^{13}$CO in the central clump (leaves 7, 8, 17 and 18, labeled in
red in Fig.~\ref{fig:pos_dendro_13co}). These are a result of the $^{13}$CO
self absorption and outflow wings, which produce an overly sub-structured tree
tracing outflowing gas at opposing velocities, and artificially separated by
the self-absorption dip. These leaves do not have the systemic velocity of
3.5~km\,s$^{-1}$ but they vary from 2.9 to 3.9~km\,s$^{-1}$ and trace the dense
outflowing gas from Flow 1.

The SE U-shaped ridge is divided into two halves in both isotopologues
(labeled in yellow in Fig.~\ref{fig:pos_dendro_13co} and
\ref{fig:pos_dendro_c18o}). The lower-left half of the U-shaped ridge is
identified with leaf 3, at slightly redshifted velocities, consistent with the
velocity map of C$^{18}$O (Fig.~\ref{fig:b59_c18o_momentum}).  The other half
of the ridge is found as a red-shifted leaf in $^{13}$CO (leaf 5), with a high
virial parameter, consistent with compressed gas from the redshifted outflow
(Flow 2).

The entire NE ridge is found as leaf 13 (labeled in green in
Fig.~\ref{fig:pos_dendro_13co} and \ref{fig:pos_dendro_c18o}), without any
sub-structure within it, and at the systemic velocity of the cloud
($\sim$3.5~km\,s$^{-1}$). This is the only leaf in B59 that appears with a
$\alpha$ below 1 in both $^{13}$CO and C$^{18}$O.

On the other hand, the western cores branch out from the main tree and form a
sub-structure of their own. They include leaves 10 and 15 (labeled in blue in
Fig.~\ref{fig:pos_dendro_13co} and \ref{fig:pos_dendro_c18o}), detected with
both molecules. The mass and virial parameter of leaf 15, in particular,
indicates that it could collapse to form future protostars, but at present it
presents no evidence for ongoing star formation. Despite having a similar mass 
to the central cores of B59, its larger size (and therefore lower densities) are 
indicative of a less compact core, which has at its disposal a smaller mass 
reservoir from where it could accrete. If it does become active, 
it will likely form no more than a single low mass 
protostar/protostellar system. Spatially nearby, there is the
blue-shifted $^{13}$CO leaf 16, and the C$^{18}$O leaf 21, both with a high
virial parameter. These two leaves lay at the edges of Cavity C, suggesting
that they are part of the material pushed out of the cavity by an
outflow. This gas-deprived cavity is now what physically separates the central
region and the western cores.

Finally, we hypothesise that the $^{13}$CO leaves 9, 11 and 14 (labeled in
purple in Fig.~\ref{fig:pos_dendro_13co}) are counterparts of the main
outflows of the region, but since they are not at very large velocity offsets
and since we cannot trace the material between these knots and the respective
driving source, it is not straightforward to understand their origin. Leaf 19,
in particular, appears isolated and is a possible outflow bullet, with the
highest virial parameter and the highest blue shifted velocity in the cloud
(at $\sim$2.2~km\,s$^{-1}$).

\subsubsection{The masses}

A brief comparison of the masses derived from the $^{13}$CO and C$^{18}$O
emission shows that the structures which are common have similar mass
estimates. C$^{18}$O recovers more mass from the NE ridge and from some of the
cores, as expected from the higher optical depth of $^{13}$CO. Nevertheless,
both lines show a total of $\sim$~10~M$_{\odot}$ in the leaves. However, they
differ in the total mass of the cluster by a factor of $\sim$2.5
($\sim$~74~M$_{\odot}$ in $^{13}$CO versus $\sim$~32~M$_{\odot}$ in
C$^{18}$O). Such a difference is mostly due to the diffuse gas traced only by
$^{13}$CO. That said, the fraction of mass in the leaves is $\sim$11~\% using
the $^{13}$CO, and $\sim$~33\% for the C$^{18}$O estimate.

Considering only the central clump of B59, our best mass estimate is retrieved
from C$^{18}$O, as estimated at the lower level of branch 45. This central
region is found to have a mass of $\sim$~16~M$_{\odot}$ within a region of
$\sim$~0.1~pc radius, with an average velocity dispersion ($\sigma_{v}$) of
0.3~km\,s$^{-1}$ (i.e. $\alpha$ = 0.7).

When using the momentum maps constructed over the entire velocity range, we
measure within a central region of $\sim$0.1~pc radii a mass of
10~M$_{\odot}$, a mean velocity of 3.5~km\,s$^{-1}$ and a mean $\sigma_{v}$ of
0.35~km\,s$^{-1}$. Assuming the uniform profile as that taken for the dendrogram
analysis, we calculate a virial parameter of 1.5. The smaller mass estimated
using the total integrated intensity is not surprising since we are now taking
a circular area around the central region of 0.1~pc radii, out of which not
all pixels are included in the calculation due to signal to noise
limitations. For the same reason, the higher velocity dispersion is also
compatible with the dendrograms, likely arising from the fact that the larger
line widths in B59 are concentrated toward the central region. This is also
clear looking at the variations of the virial parameter of branch 45
(Fig.~\ref{fig:pos_dendro_13co}, right), which starts at $\sim$2 at higher
levels, decreasing down to 0.7 at the lower level of the branch.

\end{appendix}

\end{document}